\documentclass[aps,prb,reprint,showpacs,floatfix]{revtex4-1}
\usepackage{graphicx}  %
\usepackage{dcolumn}   %
\usepackage{bm}        %
\usepackage{amssymb}   %
\usepackage{amsmath, amsthm}
\usepackage[extra]{tipa}
\usepackage{color}
\setcitestyle{square,numbers}

\emergencystretch=\maxdimen
\begin{document}
\title{Magnetic properties of alternating Hubbard ladders}
\author{Kaouther Essalah}
\affiliation{Department of Physics, Faculty of Sciences of Tunis,
University of Tunis El-Manar, Tunis 2092, Tunisia}
\author{ Ali Benali}
\affiliation{Department of Physics, Faculty of Sciences of Tunis,
University of Tunis El-Manar, Tunis 2092, Tunisia}
\author{Anas Abdelwahab}
\affiliation{Leibniz Universit\"{a}t Hannover, Institut f\"{u}r Theoretische Physik, Appelstr.~2, 30167 Hannover, Germany}
\author{Eric  Jeckelmann }
\affiliation{Leibniz Universit\"{a}t Hannover, Institut f\"{u}r Theoretische Physik, Appelstr.~2, 30167 Hannover, Germany}
\author{Richard T. Scalettar}
\affiliation{Department of Physics, One Shields Ave., University of California, Davis, California 95616, USA}

\date{\today}

\begin{abstract}

We investigate the Hubbard Hamiltonian
on ladders where the number of sites per rung alternates between two and three.
These geometries are bipartite with non-equal or equal number of sites on the two sublattices.
Thus they share a key feature of the Hubbard model 
in a class of lattices which Lieb
has shown analytically to exhibit long-range ferrimagnetic
 order, while being amenable to 
powerful numeric approaches developed for quasi-one-dimensional geometries.
The Density Matrix Renormalization Group (DMRG) method is used to obtain
the ground state properties, e.g.~excitation gaps, charge and spin densities as well as 
their correlation functions at half-filling.
We show the existence of long-range ferrimagnetic order in the one-dimensional ladder geometries.
Our work provides detailed quantitative results which complement
the general theorem of Lieb for generalized bipartite lattices.
It also addresses the issue of how the alternation between quasi-long range order
and spin liquid behavior for uniform ladders with odd and even numbers of legs
might be affected by a regular alternation pattern.
\end{abstract}


\maketitle

\section{Introduction}\label{Introduction}

Artificially constructed
quantum nanostructures of correlated electrons in quasi one (1D) 
and two (2D) dimensions can be described using lattice models such as Heisenberg and Hubbard Hamiltonians, as well as their extensions.
Novel experimental realizations of such systems include the fabrication of arrays 
of magnetic atoms on surfaces using scanning tunneling microscopy ~\cite{Tos16,Gam02,Slo17,Randall2018b,Kirk2019a}, as well as optical lattice
emulators using ultracold atoms~\cite{Bloch05,Bloch08,Greiner08}.
Bipartite lattices constitute a particularly important
class of geometries in which the
nearest-neighbor hopping (or magnetic exchange) is such that the lattice
has two subsystems $A$ and $B$ in which only $B$ sites are nearest-neighbor of $A$ sites, 
and {\it vice-versa}.
E.~Lieb proved~\cite{lieb1989two} a rigorous theorem for the Hubbard model on a bipartite 
lattice at half-filling: the ground state total spin quantum number is given by $S=(N_B-N_A)/2$, 
where $N_A$ ($N_B$) is the number of sites in the $A$ ($B$) subsystem.
Subsequently, Shen et al.~\cite{Shen1994} proved that both ferromagnetic and antiferromagnetic long-range orders coexist in the degenerate ground state with $S \neq0$, 
i.e.~the system exhibits long-range ferrimagnetic order.
Subsequent work examining Lieb's theorem 
has mostly focused on 2D lattices such as Lieb's original 
" CuO$_2$" lattice.

On the other hand,
during the past several decades, strongly correlated electron materials with 
quasi-one-dimensional (1D) ladder structures~\cite{dagotto1996surprises} 
have attracted much attention theoretically as well as experimentally. These
ladder structures reveal interesting phases including Luttinger liquids~\cite{giamarchi2004quantum}, 
Mott insulators~\cite{belitz1994anderson}, antiferromagnetism \cite{kawano1997three}, 
as well as charge density waves \cite{xu2011spin}.
Early theoretical studies of Heisenberg ladders with only nearest neighbor 
interactions and without frustration revealed an interesting effect 
by changing the number of legs~\cite{giamarchi2004quantum}. 
Geometries with an even number of legs are associated with
a singlet ground state with a spin gap 
to the lowest-lying excitations and short-range spin correlations. 
For odd numbers of legs the ground state 
has quasi-long-range antiferromagnetic order and gapless spin excitations.
Lattices with an odd number of legs are
in the universality class of the 
single leg spin-$\frac{1}{2}$ Heisenberg chain~\cite{giamarchi2004quantum}.

Hubbard ladders at half-filling (one electron per site) reveal similar 
behavior for the spin excitation modes, while the charge excitation modes remain gapped \cite{essler2005one}.
This similarity is expected since
there is a mapping between the Heisenberg model
and the spin sector of the Hubbard model
at strong values of the on-site electron repulsion $U$.
The density matrix renormalization group
(DMRG)~\cite{friedman1997density,xiang1996t,jeckelmann1998comparison} 
has proven to be an especially powerful computational tool in 
uncovering this physics.

In this article, we introduce novel ladder configurations which consist of an alternation 
of rungs with even and odd numbers of sites, and study the strong correlation physics
of the Hubbard model. 
This structure serves as an intermediary between the even rung, gapped ladder, and
the odd rung case which supports spin correlations with a power law decay.
The magnetic properties are discussed in the light of Lieb's theorem 
and the imbalanced sublattice site count induced by 
the combination of an even and odd numbers of legs.
Understanding the properties of these 
quasi-1D lattice geometries offers the opportunity to
further understand, and potentially extend, Lieb's theorem.

The paper is organized as follows. 
In the next section we discuss non-interacting nearest-neighbor tight-binding models
on alternating ladder geometries, and obtain
their band structure and density of states (DOS), which are crucial
starting points for consideration of the effects of electronic correlations. 
With this groundwork,
in section \ref{DMRG results}, we use the DMRG
method to investigate correlated Hubbard ladder systems 
with two different alternation frequencies.  One pattern
has $N_A=N_B$ and the other has $N_A \neq N_B$.  Consideration
of both cases allows us to isolate the effects of imbalanced sublattice
site counts from that of alternating number of sites per rung.
In the last section we summarize and discuss the main results of this 
work. The Appendices contain calculations for uniform 3-leg ladders,
to facilitate comparison with our studies of 
alternating ladders, and some further details for the noninteracting alternating ladders.

\section{Noninteracting alternating ladder geometries}\label{AlternatingGeometries}

In this section we derive the non-interacting
dispersion relations and DOS of two, hitherto unstudied, geometries 
with alternating rung length. 
We investigate
two alternation periods. The first has  period $d=2$ in the leg direction:
Each unit cell contains three sites in one rung and two sites in the second rung. 
We call this geometry the 3-2 ladder. It is bipartite with sublattices $A$ and $B$, 
but of the class that Lieb studied, with $N_A \neq N_B$.
The second structure has period $d=4$.  
Each unit cell contains three sites on the first and second rungs followed by two sites 
in the third and fourth rungs. We call this geometry the 3-3-2-2 ladder.
It has $N_A=N_B$.
Appendix~\ref{Homogeneous3legladder} reviews the band structure of 
uniform 3-leg ladders. 

\begin{figure}
    \includegraphics[width=0.4\textwidth]{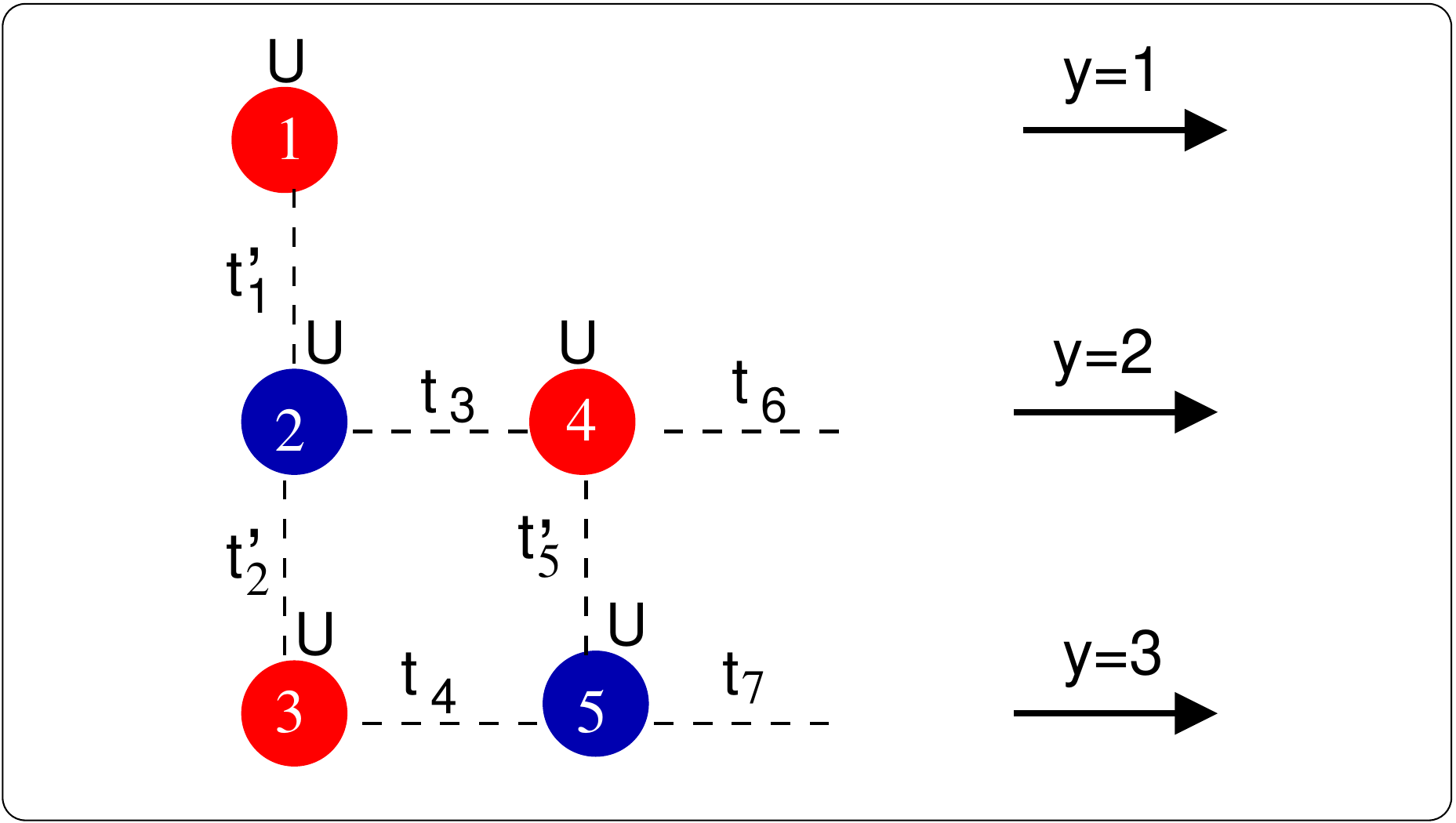}
    \caption{\label{geometry2-3} 3-2 ladder geometry with  alternating rung site numbers. 
    The red and blue circles indicates the sites belonging  to the sublattices A and B, respectively.
    Dashed lines label the non-zero hopping terms.
    }
\end{figure}

The  nearest-neighbor tight-binding model is described by the Hamiltonian
\begin{eqnarray}\label{3legHubbardU0}
H_{0} &=&  -  \sum_{x, y, \sigma} t_{x,y} \left( c^{\dag}_{x+1, y, \sigma} c^{\phantom{{\dag}}}_{x, y, \sigma}+ c^{\dag}_{x, y, \sigma} c^{\phantom{{\dag}}}_{x+1, y, \sigma}\right)
\nonumber \\
&-&  \sum_{x, y, \sigma} t^{\, \prime}_{x,y} \left( c^{\dag}_{x, y+1, \sigma} c^{\phantom{{\dag}}}_{x, y, \sigma} + c^{\dag}_{x, y, \sigma} c^{\phantom{{\dag}}}_{x, y+1, \sigma} \right) .
\end{eqnarray}
$c^{\dag}_{x, y, \sigma} \left( c^{\phantom{{\dag}}}_{x, y, \sigma} \right)$ denotes the creation (annihilation) operators for an electron with spin $\sigma$ on the site with coordinates $\left( x, y \right)$ where $y=1,2,3$ denotes the Hubbard leg and $x = 1, \dots,L_x$ refers to the rung index.
The parameters $t_{x,y}$ and $t^{\, \prime}_{x,y}$ are hopping amplitudes along and between the chains, respectively. 
They depend on the ladder geometry considered.

\begin{figure}
    \includegraphics[width=0.48\textwidth]{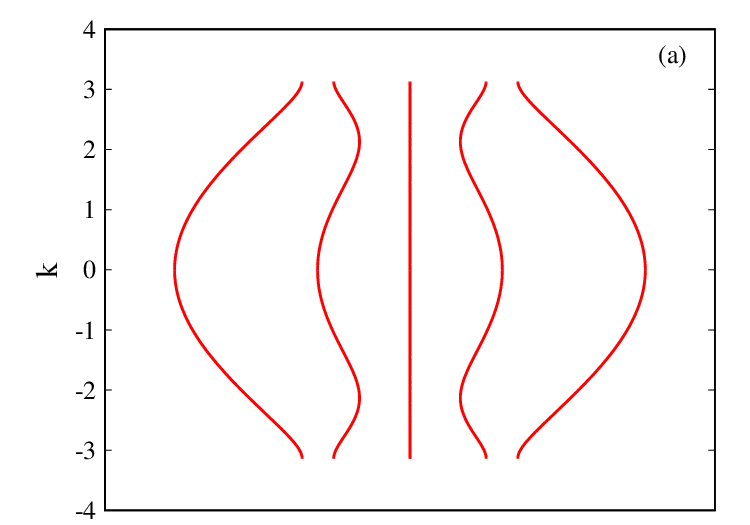}
    \includegraphics[width=0.48\textwidth]{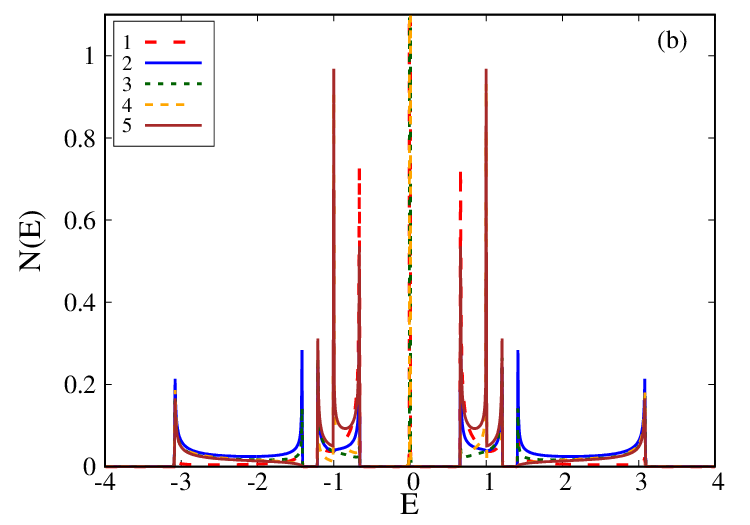}    
    \caption{\label{3-2-non-interacting-rest} (a) Band structure for the 3-2 alternating 
    ladder in the noninteracting case for uniform hopping terms $t_i=t=1$. (b) Corresponding local density of states~(\ref{LDOS}) for each site $j=1,\dots,5$
    of the unit cell shown in Fig.~\ref{geometry2-3}. 
     }
\end{figure}

\subsection{The alternating 3-2 ladder}

The 3-2 ladder system is sketched in Fig.~\ref{geometry2-3}. 
Defining $L_x$ as the total number of rungs,
the total numbers of sites on the two sublattices are unequal, with
$N_A=3L_x/2$ and $N_B=L_x$.

$H_0$ can be written as a sum of commuting operators $H_k$ acting only
on the Bloch states with wave number $k$ in the first Brillouin zone (see Appendix~\ref{Homogeneous3legladder} to get the concrete form of wave number $k$). In the present case, we have
five sites
per unit cell, labeled $j=1,\dots,5$ in Fig.~\ref{geometry2-3}. 
$H_k$ is the $5 \times 5$ matrix,
\begin{equation}
H_{k}= 
\begin{pmatrix}
   0 &  -t_1   &  0  &   0   & 0 \\
  -t_1  &   0 &-t_2  &-t_3-\widetilde t_6 & 0 \\
   0  &  -t_2 &   0    &  0 & -t_4-\widetilde t_7 \\
   0   & -t_3-\widetilde t^{\ast}_6&  0 &  0 & -t_5\\
   0   &  0  &  -t_4-\widetilde t^{\ast}_7& -t_5  & 0
\end{pmatrix}.
\label{32MomentumSpace}
\end{equation}
Here $\widetilde t_j =t_j\exp(ik)$.

\begin{figure}
	\includegraphics[width=0.4\textwidth]{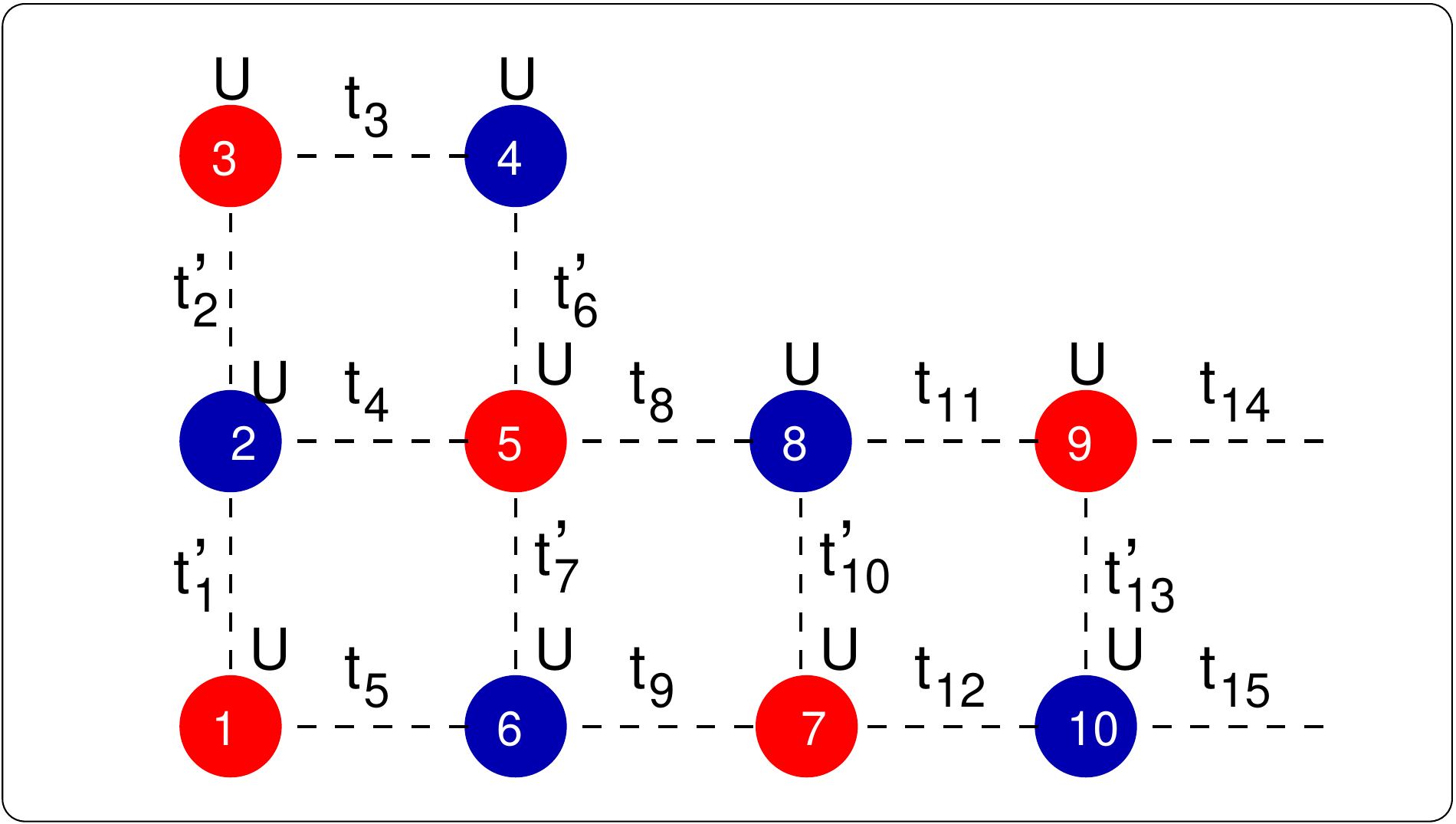}
	\caption{\label{4-6geometry} Geometry of the alternating 3-3-2-2 ladder lattice. The red and blue circles indicate the  sites belonging to the $A$ and $B$ sublattice, respectively.
	Dashed lines show the non-zero hopping terms.
	}
\end{figure}

We will investigate the system with equal hopping parameters  
$t_i=t$ for $i=1, ..., 7$, since this simple choice already contains
several novel features.
The diagonalization of 
$H_{k}$ 
leads to the five energy bands $E_{k,b}$ ($b=1,\dots,5$) shown 
in Fig.~\ref{3-2-non-interacting-rest}(a). One of these bands is flat and is located at zero energy 
(the Fermi level for half-filling), in accordance with 
Lieb's theorem~\cite{lieb1989two}.
For such a flat band, 
there are gapless single-electron excitations but the system is not metallic in the
sense that the effective mass of the low-lying
excitations, which is proportional
to $1/t$ for a linear chain, is infinite.
Figure~\ref{3-2-non-interacting-rest}(b) shows the corresponding local DOS
\begin{equation}
 N(E,j) = \frac{1}{N_c} \sum_{{k,b}} \vert\psi_{k,b}(j)\vert^{2} \delta{(E-E_{k,b})}
 \label{LDOS}
\end{equation}
where  $\psi_{k,b}(j)$ represent the eigenvectors of Eq.~(\ref{32MomentumSpace}) corresponding to the 
eigenenergy $E_{k, b}$, $j=1,\dots,5$ are the sites of the unit cell.  $N_c=L_x/2$ 
is the number of unit cells (or equivalently the number of wave vectors $k$ in the first Brillouin zone). 
Note that all DOS distributions in this paper are normalized so that the integral over all energies $E$ is equal to 1.
To draw the DOS we substitute a Lorentzian function of width $\eta=0.01 t$ for the  Dirac $\delta$-function.
The singularity at 
$E=0$ in the local DOS is due  to the flat band. 
Figure~\ref{3-2-non-interacting-rest}(b)
reveals that this flat band is located on the $A$ subsystem, ie. sites 1, 3 and 4 in the unit cell in
Fig.~\ref{geometry2-3}.

\subsection{Alternating 3-3-2-2 ladder}\label{Alternating rung geometry; 3-3-2-2}

The real space 3-3-2-2 ladder geometry is displayed in Fig.~\ref{4-6geometry}.
The lattice is bipartite, with an equal number of sites in
the sublattices $A$ and $B$. 
Similarly to 
the 3-2 ladder geometry, $H_0$ can be written as a sum of commuting 
operators $H_k$ acting only
on the Bloch states with wave number $k$. 
Here 
there are 10 sites per unit cell.
Its single-particle matrix representation $H_k$ 
is the $10 \times 10$ matrix given in Appendix~\ref{app:matrix}.

\begin{figure}
 \includegraphics[width=0.48\textwidth]
    {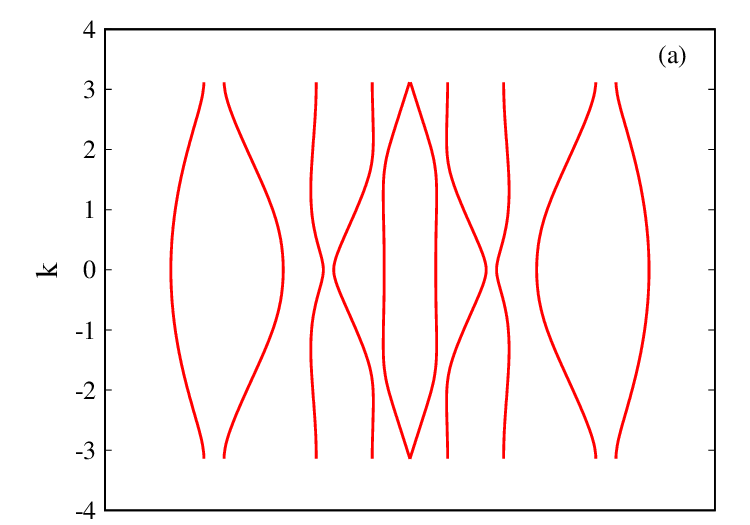}
 \includegraphics[width=0.48\textwidth]
    {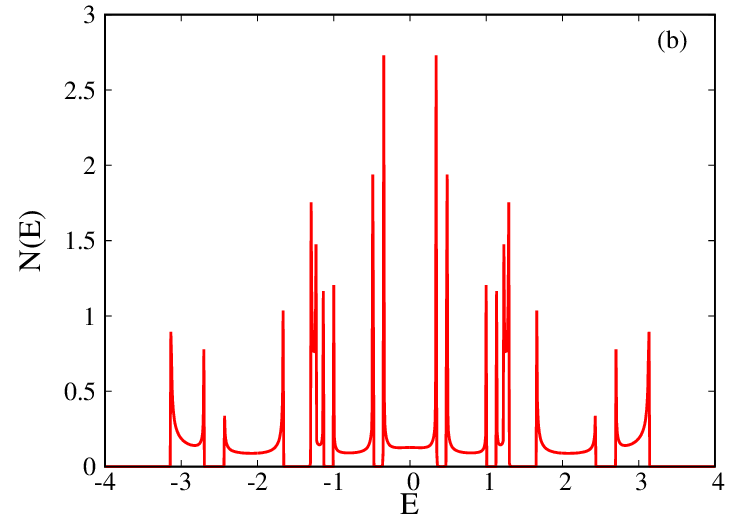}
    \caption{\label{Band-structure-10} (a) Band structure and (b) DOS, Eq.\ref{DOS}, of the
noninteracting alternating 3-3-2-2 ladder. Compared to the 3-2 geometry in Fig.\ref{3-2-non-interacting-rest} the key difference is the absence of the flat band at $E = 0$.
    }
\end{figure}

The diagonalization of the matrices 

$H_k$ 
leads to the ten energy bands $E_{k,b}$ shown in Fig.~\ref{Band-structure-10}(a). Since $N_A=N_B$,  we do not observe a flat band.
The system is metallic at half filling as the Fermi 
energy $E_F = 0$ lies 
at a point where two bands touch.
This is also visible in the total DOS 
\begin{equation}
\label{DOS}
 N(E) = \frac{1}{N_s} \sum_{{k,b}} \delta{(E-E_{k,b})} 
\end{equation}
with division by the number of sites $N_s$
providing the previously described normalization convention for $N(E)$.
This DOS is plotted in 
Fig.~\ref{Band-structure-10}(b), which shows a finite density but no concentration of spectral weight at the Fermi level, in contrast to the peak seen 
in Fig.~\ref{3-2-non-interacting-rest}(b).

Although the 3-2 and 3-3-2-2 geometries share an alternation of number of sites
in different rungs, they differ significantly
due to the presence of a flat band in the 3-2 case.  
We will analyze in the next section how this difference is reflected in the 
properties of ladders with the on-site interaction turned on.

\section{DMRG results for alternating ladders}\label{DMRG results}

In this section we analyze and compare the properties of the 3-2 and 3-3-2-2 ladder geometries in the presence of the Hubbard interaction using the finite-size DMRG method.
 The full Hamiltonian is
\begin{equation}
H=H_0  + U \sum_{x, y} n_{x, y ,\uparrow} n_{x, y, \downarrow}
\label{eq:Hfull}
\end{equation}
with $n_{x, y , \sigma} = c^{\dag}_{x, y , \sigma} c^{\phantom{{\dag}}}_{x, y , \sigma}$.

DMRG is widely considered to be the most powerful numerical method for quasi-1D correlated electron systems~\cite{white1993density}. The details of this method have been reviewed in Ref.~\cite{schollwock2005density}.
In our DMRG calculations, open boundary conditions are applied in both $x$- and $y$-directions.
Our program uses the conservation of the particle numbers $N_{\sigma}$ but not the SU(2) spin symmetry.
The number $m$ of density-matrix eigenstates in the renormalization procedure is increased progressively until it reaches $m=2500$ in the last sweep.
 The total number of used sweeps is up to $13$. The discarded density-matrix weight (truncation error) varies from $10^{-5}$ to less than $10^{-9}$.
We extrapolate the ground state energy to the vanishing discarded weight limit~\cite{bonvca2000stripes} and estimate the error from the difference between the extrapolated energy and the energy in the last sweep. The finite-size error is obtained by varying the system length and extrapolating to vanishing ratio $1/L_x$.
In our calculations the ladder length $L_x$  is varied up to 200.
 We have implemented the one-dimensional 
DMRG path through all lattice sites of the alternating ladder geometries so that
the sites are ordered as in the standard approach for 
homogeneous ladders~\cite{schollwock2005density}.
The ground state energy, spin, charge and single-particle gaps, 
the pair-binding energies, correlation functions, 
as well as spin and charge densities are investigated for a wide range of parameters.
We use homogeneous hopping parameters $t_i=1$ everywhere.

\subsection{3-2 alternating rung geometry}	

\begin{figure}
    \includegraphics[width=0.42\textwidth,angle=-90]{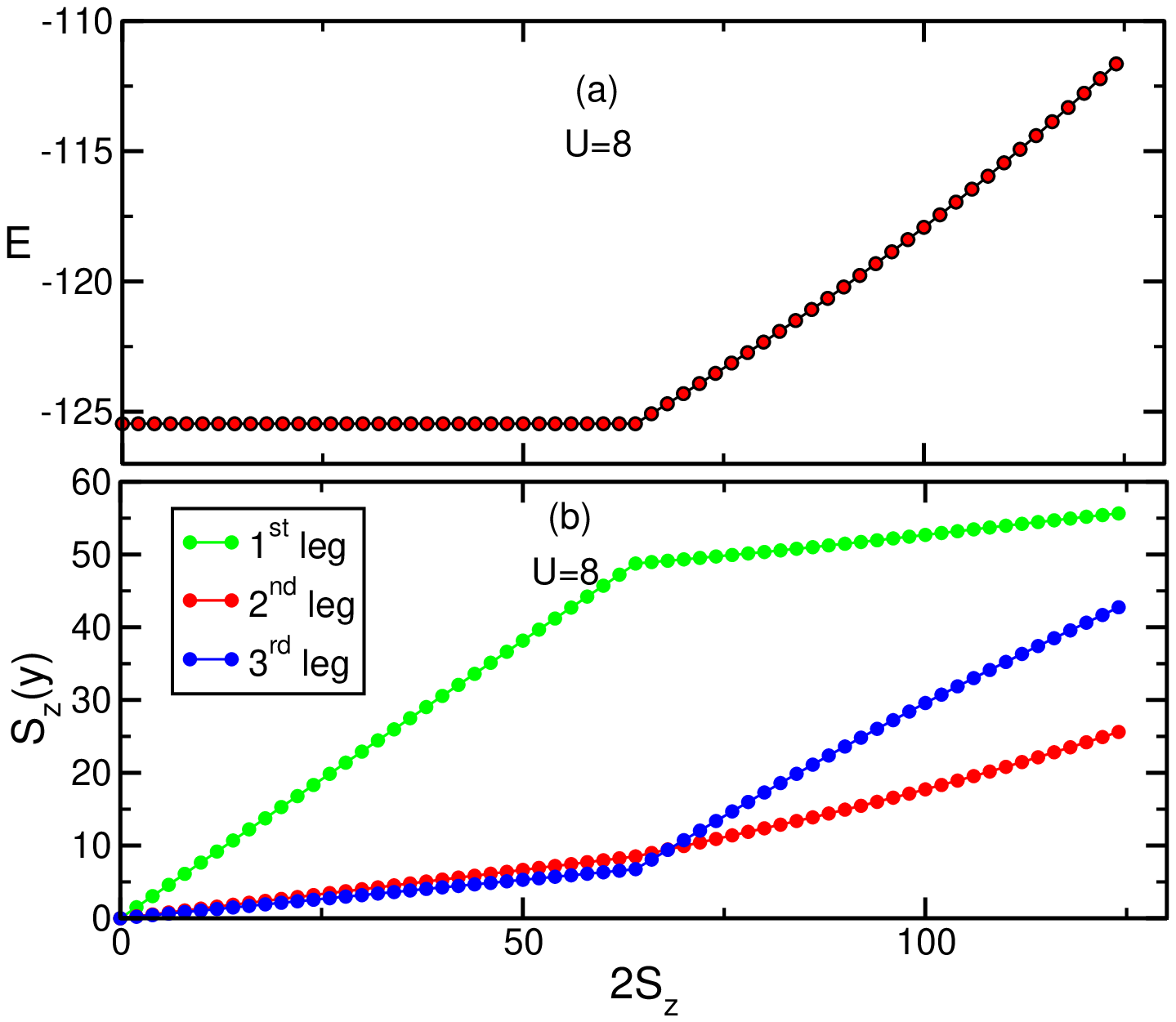}
	\caption{\label{3-2-GS-total-Sz-U8} (a) Energy of the lowest eigenstate
	and (b) its total spin density  $S_z(y)$ on each leg of the 
	3-2 ladder geometry
	as a function of the $z$-projection of the eigenstate total spin $S_z$ for $L_x=128$, half filling and $U=8$. Note that the energy $E$ is an even function of $S_z$ while the densities are odd functions of $S_z$.} 
	
\end{figure}

According to Ref.~\cite{lieb1989two}, we expect the ground state at 
half filling to be ferromagnetic, with 
total spin
 $S=(N_A  - N_B)/2 = L_x  / 4$ for any finite value of the 
on-site interaction $U>0$.
We have first computed the eigenstate with the lowest energy as a 
function of the $z$-projection of the 
total spin $S_z=(N_{\uparrow}-N_{\downarrow})/2$.
Results for the corresponding eigenenergies are shown in Fig.~\ref{3-2-GS-total-Sz-U8}(a) for $U=8$ and $L_x=128$.
The ground state is degenerate for all $\vert S_z \vert  \leq S = 32$ and thus ferromagnetic,
in agreement with Lieb's prediction.
For $\vert S_z \vert  > 32$ the eigenstates are excited states.
As $S$ increases proportionally to the ladder length $L_x$, the ground state is macroscopically degenerate in the thermodynamic limit.

The DMRG method can be used to compute the charge and spin density profiles, 
defined by
\begin{equation} \label{spin density}
N (x,y)= \langle n_{x,y,\uparrow}+n_{x,y,\downarrow} \rangle
\end{equation} \label{charge density}
\vspace{-0.5cm}
\begin{equation}
\label{spin_density}
S_z(x,y)= \langle n_{x,y,\uparrow}-n_{x,y,\downarrow} \rangle 
\end{equation}
where $ \langle  \dots  \rangle$ denotes the expectation value in the lowest eigenstate for 
a given number $N_{\sigma}$ of electrons of each spin in the system.  
Due to the particle-hole symmetry of the Hamiltonian~(\ref{eq:Hfull}), 
the charge density is distributed homogeneously $N(x,y)=1$
at half-filling, despite the nominal inequivalence of different sites. This is even 
the case for arbitrary, unequal values of the nearest-neighbor hopping terms $t_i$. 

The behavior of the total spin density for each leg
\begin{equation}
S_z(y) = \sum_x  S_z(x,y)
\end{equation}
is, however, non-trivial. $S_z(y)$
is depicted in Fig.~\ref{3-2-GS-total-Sz-U8}(b) as a function of the $z$-projection 
of the total spin, for  $L_x=128$ and $U=8$.
We see that the spin density on each leg increases linearly for the values 
$\vert S_z \vert \leq L_x/4$,
corresponding to the degenerate ground state in 
Fig.~\ref{3-2-GS-total-Sz-U8}(a). 
For higher
$S_z$, corresponding to excited states, 
the increase continues linearly but with a different slope.
Most of the spin density
is concentrated on the first leg
for $0 < \vert S_z \vert \leq L_x /4$. Thus the ferromagnetic state is
due to the unpaired electrons localized on this first leg. We remark that the macroscopic degeneracy in the thermodynamic limit is a
consequence of Lieb's theorem in which our numerical results agree with it.

Additional information is provided by the gaps to the various 
excitations.
The spin gap
corresponds to the lowest excitation energy from the ground state 
with $S_z=0$ to an excited state in which the up and down fermion numbers 
differ by one ($S_z=1$).   That is,
\begin{equation}
E_s = E(N_{\sigma} + 1,N_{\sigma} - 1) - E(N_{\sigma},N_{\sigma})
\end{equation}
where $E(N_{\uparrow},N_{\downarrow})$ is the ground-state energy for a Hubbard lattice with $N_{\sigma}$ electrons
of each spin $\sigma$.
Experimentally, the spin gap 
$E_s$ can be determined from the dynamical spin structure factor measured using inelastic neutron scattering \cite{furrer2009neutron}. 

\begin{figure}
	\includegraphics[width=0.4\textwidth,angle=-90]{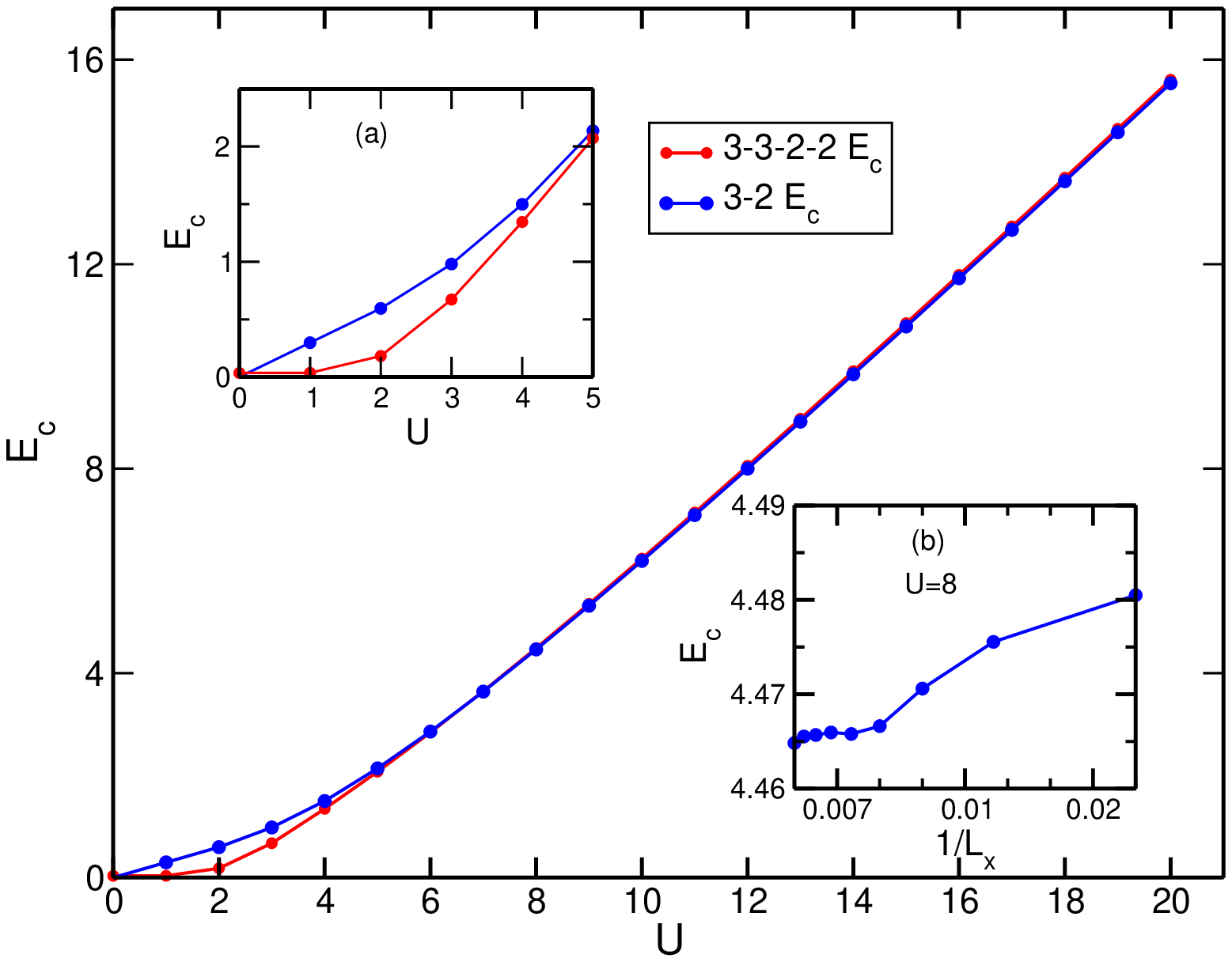}
	\caption{\label{6-4-3-2-Ec-U} Charge gap for the 3-2 and 3-3-2-2 ladders as a function of the interaction strength $U$ for $L_x = 128$ at
half-filling. The inset (a) shows the small coupling region on a smaller scale. The inset (b) shows the charge gap $E_c$ of the 3-2 ladder geometry at half-filling as a function of the inverse ladder length $1/L_x$ for $U=8$.}
\end{figure}

The charge gap
is the lowest excitation energy from the $N_e$-particle ground state 
to the ($N_e\pm2$)-particle ground states with the same $S_z$. Its experimental
value is deducible from, for example, the gap in the dynamical charge structure 
factor measured using electron-energy-loss spectroscopy \cite{egerton2011electron}.
It is defined as~\cite{noack1995doped}
\begin{eqnarray}
E_c &=& \frac{1}{2}\left [E(N_{\uparrow} + 1,N_{\downarrow} + 1) + E(N_{\uparrow} - 1,N_{\downarrow}-1) \right]\nonumber\\
&-& E(N_{\uparrow},N_{\downarrow}).
\label{charge-gap-equation}
\end{eqnarray}

Finally, the single-particle gap is the lowest excitation energy $E_{p}$ seen in the single-particle 
spectral function, which can be measured in experiments such as 
Angle Resolved Photo-emission Spectroscopy (ARPES) \cite{zhang2012photoemission}
\begin{eqnarray}
 E_{p} &=& E(N_{\uparrow} + 1,N_{\downarrow}) +  E(N_{\uparrow}-1,N_{\downarrow}  )\nonumber\\
 &-& 2E(N_{\uparrow},N_{\downarrow}).
 \label{single-particle-gap-equation}
\end{eqnarray}
Consequently, $E_{p}$ is the gap due to the excitation of a single-electron 
(with both charge and spin features) from the highest level below the Fermi 
level to the lowest level above the Fermi level. 

The three gaps vanish in the half-filled noninteracting 3-2 ladder geometry. 
The spin gap remains small for a coupling $U>0$ at finite system size $L_x$ and  extrapolates  to zero within numerical accuracy. This is consistent with the degeneracy of the ferromagnetic ground state.
An interaction $U>0$ generates a gap to the lowest  
charge excitations.
The charge gap $E_c$, depicted in Fig.~\ref{6-4-3-2-Ec-U} for $L_x=128$, evolves monotonically with increasing $U$.
$E_c$ extrapolates to finite values for $1/L_x\rightarrow 0$
(see the inset (b) of Fig.~\ref{6-4-3-2-Ec-U}) 
 in agreement with Lieb's prediction~\cite{lieb1989two}.  
The small finite size slope reflects the low velocity (almost zero) of the lowest excitations in this system. 
As for the homogeneous ladder,
the charge gap is roughly linear in $U$ at strong coupling.
 Finally, the single-particle gap $E_p$ extrapolates to the same finite value as the charge gap $E_c$ for $1/L_x\rightarrow 0$ with very small finite size effects. This behavior of excitation gaps is characteristic for the insulating ferromagnetic phase of the 3-2 geometry

\begin{figure}
   \includegraphics[width=0.4\textwidth,angle=-90]{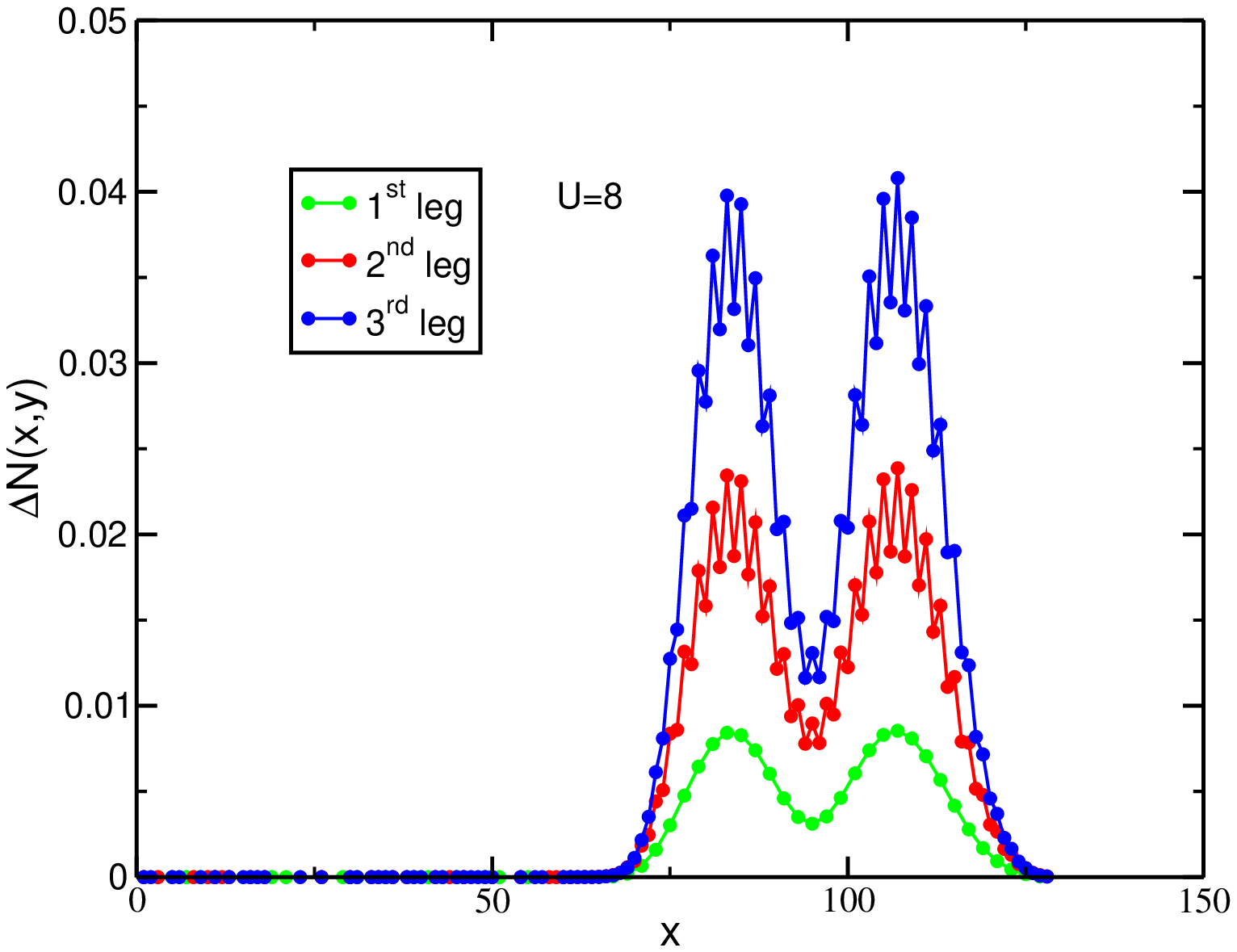}
    \caption{\label{3-2-chag-density} 
    Change
    $\Delta N(x,y)$ in the charge density distribution $N(x,y)$ on the three legs ($y=1,2,3$) when two electrons are added to a  half-filled 3-2 ladder with $U=8$, $L_x=128$, and $S_z=32$.
    }
\end{figure}

Although at half-filling
the particle-hole symmetry implies that the charge density is 
distributed homogeneously between the three legs, 
this symmetry is lost away from half filling.   
Using DMRG we calculate the change $\Delta N(x,y)$ in the charge density distribution $N(x,y)$ when two electrons are added to a half-filled ladder.
Figure \ref{3-2-chag-density} shows the results 
for each leg of the half-filled 3-2 ladder at maximal $S_z$. 
The added charges are mainly on the third and 
second legs. 
One sees the presence of a double density peak that is typical for two independent particles in a box. This suggests that the two added electrons do not bind
in this system in agreement with the vanishing
of the pair binding energy in the thermodynamic limit
($E_c=E_p$).
Note that the asymmetric distribution $\Delta N(x,y)$ (i.e.~added charge not centered at $x=L_x/2=64$) 
is due to a poor DMRG convergence and reveals that the charge excitation band width
is narrow as already suggested by the small finite-size 
corrections to the excitation gaps in Fig.~\ref{6-4-3-2-Ec-U}, inset(b).

According to Shen et al.~\cite{Shen1994} a half-filled Hubbard model on a bipartite lattice with $N_A \neq N_B$
should exhibit a ferrimagnetic long-range order. 
More precisely,
spins on the same sublattice should be ferromagnetically ordered while spin pairs on different sublattices should be antiferromagnetically ordered. This exact result applies to the 3-2 ladder geometry. Thus we should be able to observe this one-dimensional long-range order although it breaks the continuous SU(2) symmetry of the spin sector in the Hubbard model. 

For this purpose we now investigate the ground-state spin correlations. 
The transverse spin correlation function between a site $x_0,y=1$ 
and the other sites $x_0+x,y$ is defined by
\begin{equation}
C^{+-}_{y}(x) = \big\langle  \, c^{\dag}_{x_0,1,\uparrow} 
\, c^{\phantom{\dag}}_{x_0,1,\downarrow} \,
 c^{\dag}_{x_0+x,y,\downarrow} \, c^{\phantom{\dag}}_{x_0+x,y,\uparrow} \,
\big\rangle .
\label{Spin_correlation_function}
\end{equation}
These correlations are calculated using the DMRG method with the reference 
site $x_0$ located in the center of the chain, i.e. $x_0=L_{x}/2$, 
in order to minimize boundary effects.

\begin{figure}
    \includegraphics[width=0.4\textwidth,angle=-90]{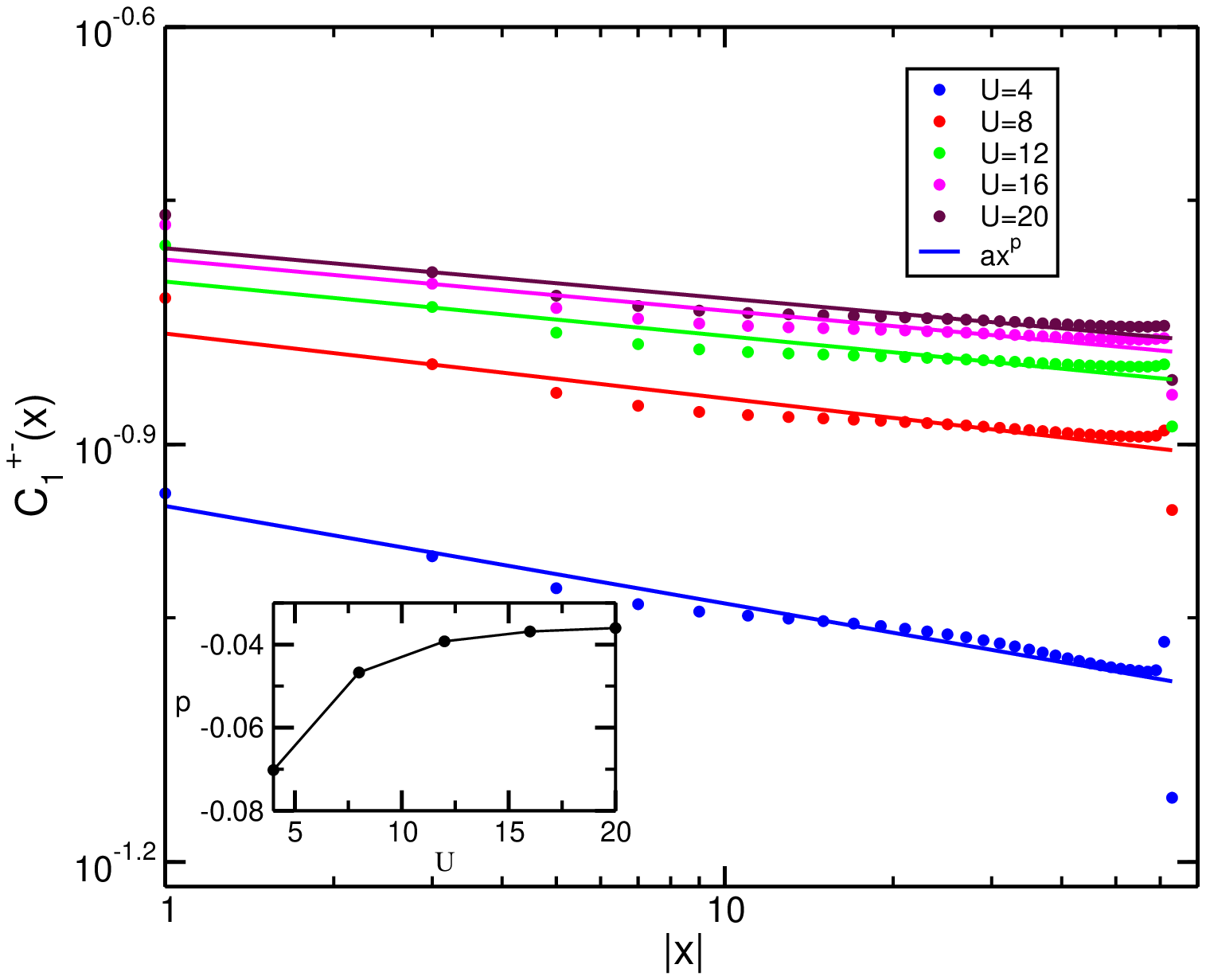}
    \caption{\label{3-2-spin-correlations-Sz-0} Transverse spin correlation function $C^{+-}_1(x)$  in the
    $S_z=0$ ground state of the the 3-2 alternating ladder. Dots show the DMRG results calculated at half-filling for $L_x=128$ and various values of $U$. The straight lines represent the fitting function $ax^p$. 
    The inset shows the exponent $p$ as a function of $U$.}
\end{figure}

According to Ref.~\cite{Shen1994} this correlation function should reveal the ferrimagnetic long-range order
between the spins in the $S_z=0$ ground state.
Figure~\ref{3-2-spin-correlations-Sz-0}
shows DMRG results for the first leg, $C^{+-}_1(x)$, for a half-filled 3-2 ladder
with $S_z=0$ and several values of the interaction $U$. 
The correlations decay very slowly with increasing distance $x$ as shown by the log-log scale.
A fit to a power-law function yields very small exponents 
$\vert p\vert \alt 0.07$ as seen in the inset of 
Fig.~\ref{3-2-spin-correlations-Sz-0}.
Antiferromagnetic correlations decrease as $1/x$ in the antiferromagnetic
isotropic Heisenberg chain. 
Fig.~\ref{3-2-spin-correlations-Sz-0} indicates
our 3-2 ladder system corresponds to the
ferromagnetic
isotropic Heisenberg chain. In both these models the ground state is
macroscopically degenerate and has true long-range ferromagnetic order. If you approach the isotropic point of the Heisenberg chain from its Luttinger liquid phase, ferromagnetic correlations decay as $x^{-p}$ with an exponent $p$ that vanishes at the
isotropic point \cite{giamarchi2004quantum}.
While the homogeneous 3-leg ladder is in the universality
class of the
antiferromagnetic Heisenberg chain, the novel 3-2 ladder considered here seems to be in the
class of the ferromagnetic chain.

The correlation functions for
the second and third legs, $C^{+-}_y(x)$ for $y=2,3$,
reveal long-range antiferromagnetic correlations (not shown). Thus our results for the transverse correlation functions $C^{+-}_y(x)$ in the $S_z=0$ ground state 
agree with the exact results~\cite{Shen1994}.

\begin{figure}
    \includegraphics[width=0.4\textwidth,angle=-90]{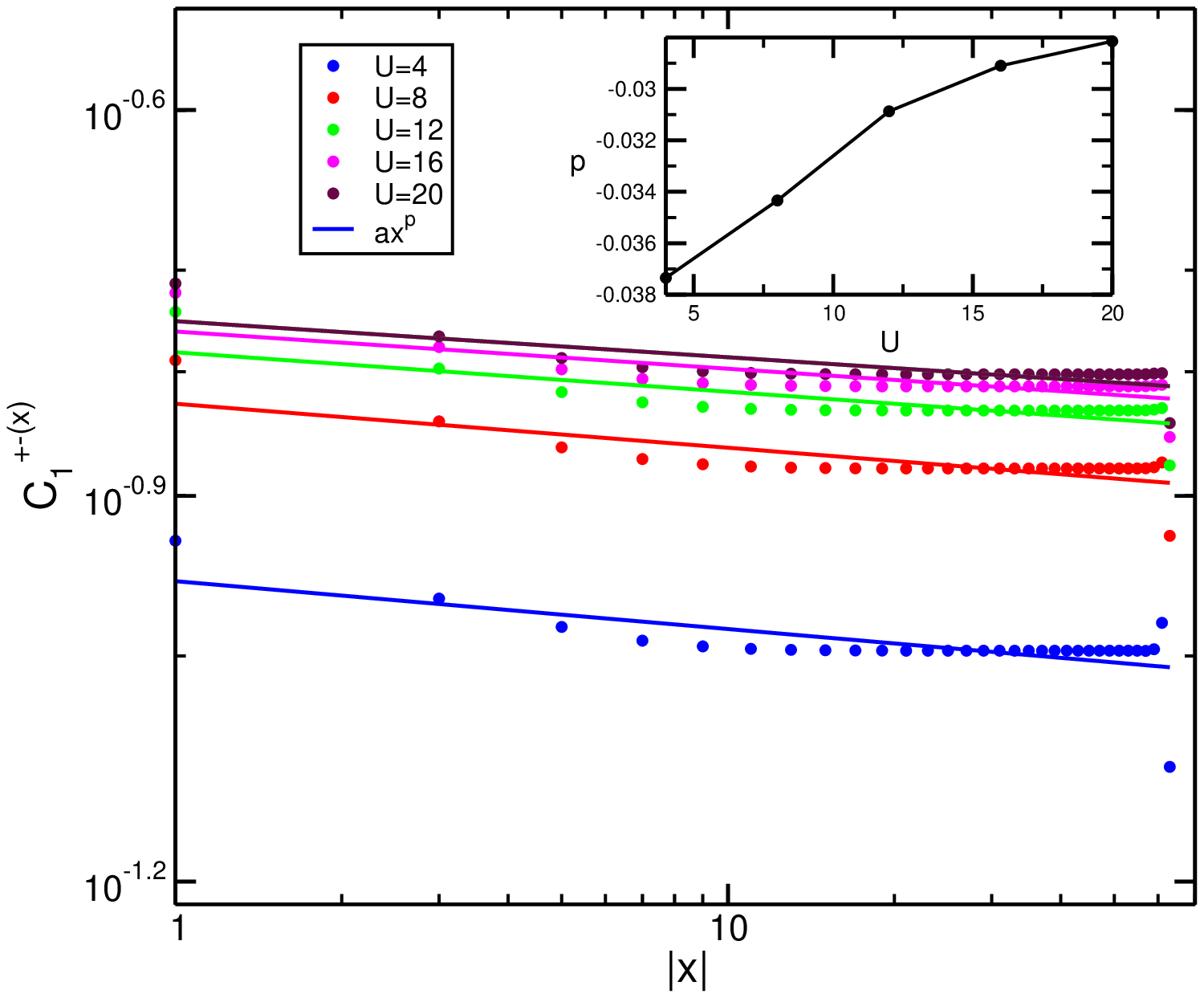}
    \caption{\label{3-2-spin-correlations-Sz-64} Transverse spin correlation function $C^{+-}_1(x)$
    in the
    $S_z=S=32$ ground state of the the 3-2 alternating ladder. Dots show DMRG results at half-filling for $L_x=128$ and various values of $U$. The straight lines represent the fitting function $ax^p$.
    The inset shows the exponent $p$ as a function of $U$.}
\end{figure}

To the best of our knowledge there is no concrete prediction for the correlation functions $C^{+-}_y(x)$ in the ground states with  $S_z\neq 0$. 
However, the exact results of Ref.~\cite{Shen1994}
imply that the SU(2) invariant correlation function
\begin{equation}
\label{eq:symcorr}
C_{y}(x) = \big\langle  \, \vec{S}_{x_0,y} \, \cdot \,
 \vec{S}_{x_0+x,y} 
\big\rangle 
\end{equation}
must also exhibit long-range ferrimagnetic order for all ground states
$-S \leq S_z \leq S$.
Thus $C^{+-}_y(x)$, and the complementary longitudinal correlation function
\begin{equation}
C^{zz}_{y}(x) = \langle (n_{x_0,1,\uparrow}-n_{x_0,1,\downarrow})
( n_{x_0+x,y,\uparrow}-n_{x_0+x,y,\downarrow}) \rangle \, ,
\label{3-2-sz-correlations-2} 
\end{equation}
must display long-range ferrimagnetic order.

Our DMRG results reveal that the spin correlation $C^{+-}_y(x)$ decay very slowly 
for ground states with spin $S_z \neq 0$.
Figure \ref{3-2-spin-correlations-Sz-64}
shows the correlation function $C^{+-}_1(x)$ on the first leg for the ground state with maximal spin $S_z=32$. The exponents of power-law fits, $\vert p\vert \alt 0.04$,
are even smaller than for
the $S_z=0$ ground state.
Again, this is compatible with long-range ferromagnetic order in the thermodynamic limit.

\begin{figure}
    \includegraphics[width=0.4\textwidth,angle=-90]{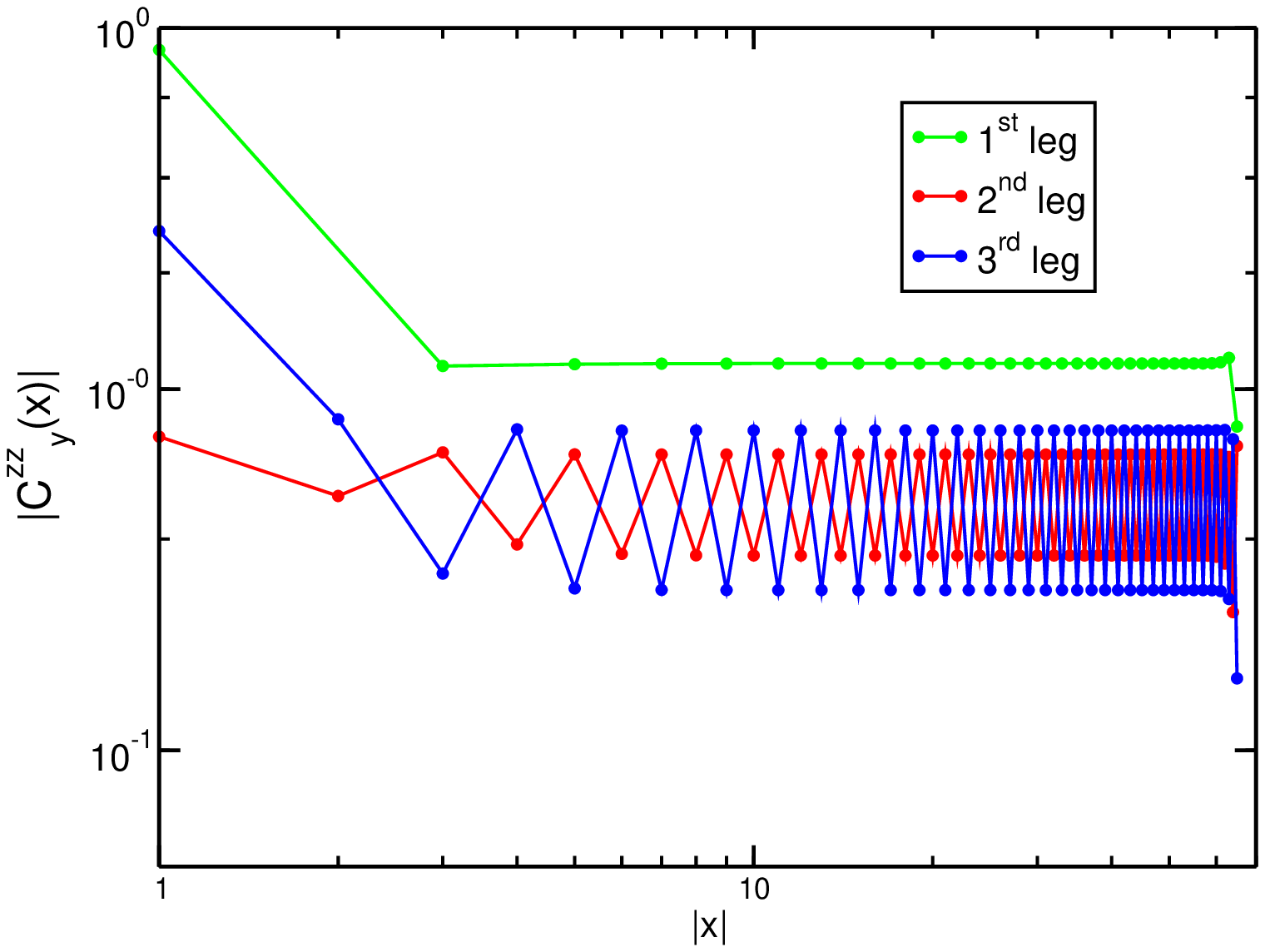}
    \caption{\label{3-2-sz-Corre-WithoutDensitySubtractionA} Longitudinal spin correlation function $C^{zz}_y(x)$ in the
    $S_z=S=32$ ground state of the the 3-2 alternating ladder. Dots show DMRG results at half-filling for $L_x=128$ and $U=8$. 
    }
\end{figure}

\begin{figure}
    \includegraphics[width=0.4\textwidth,angle=-90]{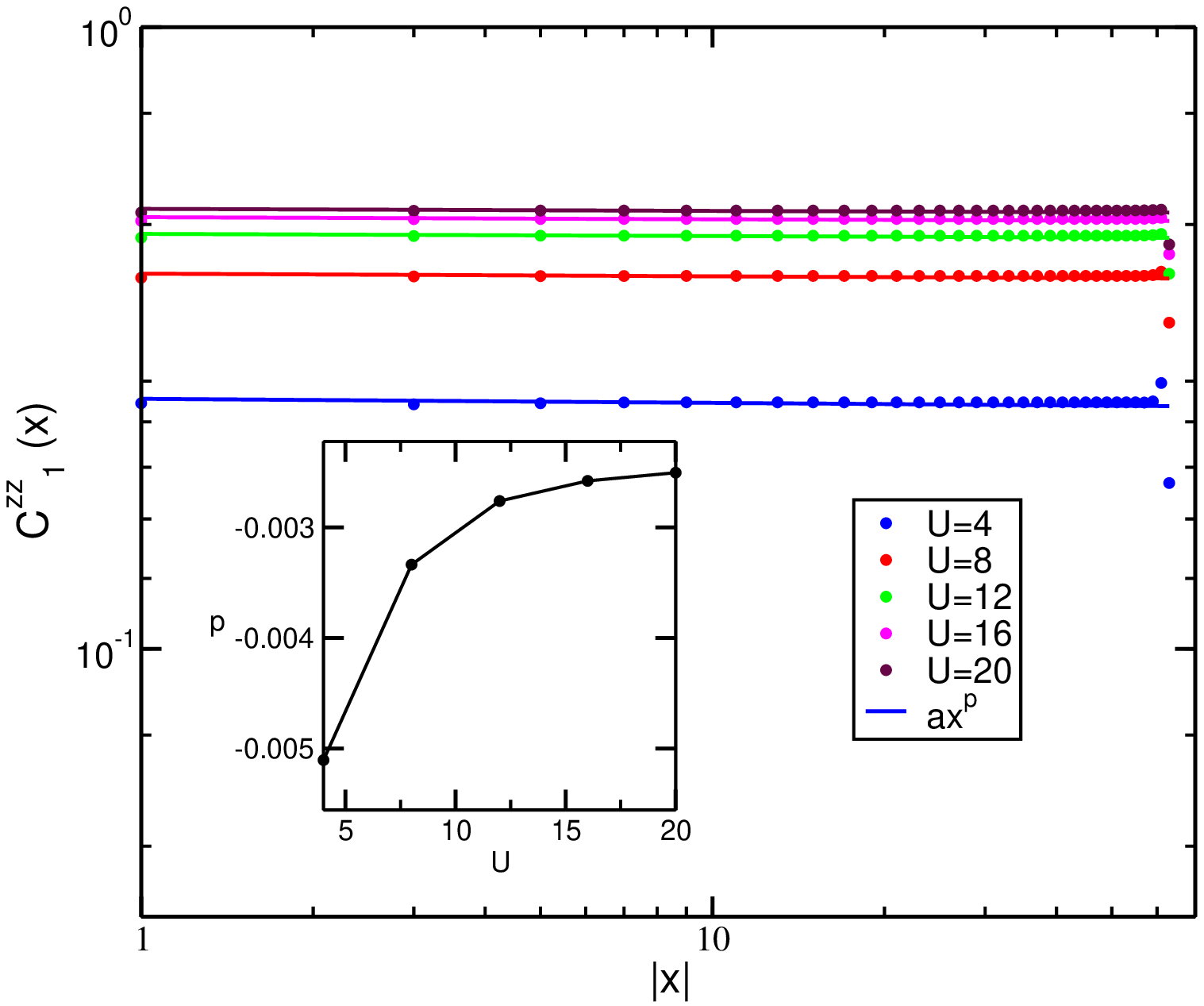}
    \caption{\label{3-2-sz-Correlations-WithoutDensitySubtraction} Longitudinal spin correlation function $C^{zz}_1(x)$ in the
    $S_z=S=32$ ground state of the the 3-2 alternating ladder. Dots show DMRG results at half-filling for $L_x=128$ and various values of $U$. 
The straight lines represent the fitting function $ax^p$.
    The inset shows the exponent $p$ as a function of $U$.
    }
\end{figure}

Turning to the longitudinal correlation function $C^{zz}_{y}(x)$,
we find that it decays rapidly for the $S_z=0$ ground state. For the ground states with $S_z\neq 0$, however,
$C^{zz}_{y}(x)$ can also reveal the long-range ferrimagnetic order. 
Figure \ref{3-2-sz-Corre-WithoutDensitySubtractionA} plots $C^{zz}_y(x)$ for all legs $y=1,2,3$ at fixed $U$
for the ground state with the maximal spin $S_z=S$.
Figure \ref{3-2-sz-Correlations-WithoutDensitySubtraction}
shows the behavior of these correlations on the first leg only, varying $U$.
Clearly, the ferromagnetic correlations do not decay with distance. A power-law fit yields exponents $\vert p\vert \alt 0.005$, as seen in the inset of this figure.

\begin{figure}[b]
    \includegraphics[width=0.4 \textwidth]{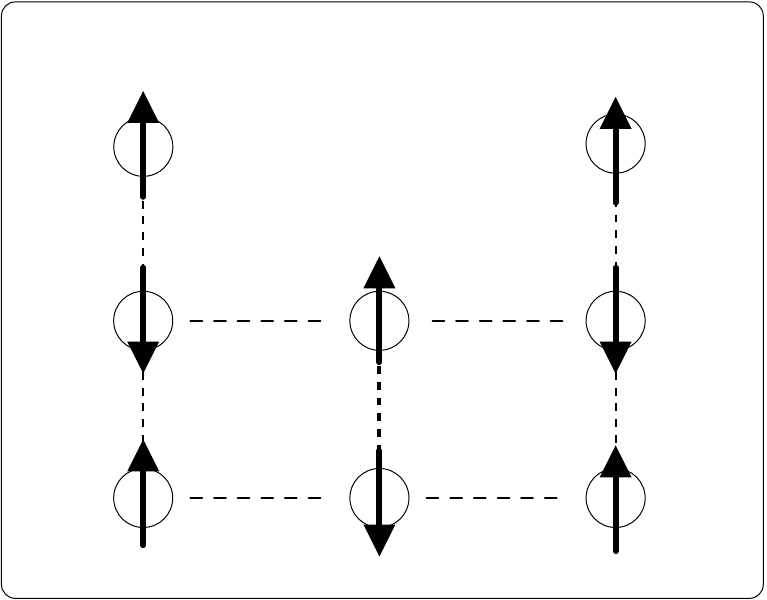}
    \caption{\label{3-2-spin-bond-order} 
    Schematic illustration of the ferrimagnetic spin ordering in the 3-2 alternating ladder.
    }
\end{figure}

The ferrimagnetic long-range order is not visible in the spin density (Eq.~\ref{spin_density}) of the $S_z=0$ ground state
because the spin flip symmetry imposes $S_z(x,y)=0$.
However, the existence of this ordering can be seen in the spin density of the other ground states, in particular for the maximal spin $S_z=S$. The sign of 
these spin densities alternates between nearest-neighbor sites as sketched in Fig.~\ref{3-2-spin-bond-order}.

The ferromagnetic order parameter is the magnetization pro site $\vert N_B - N_A \vert/(N_A+N_B)$, which is independent of $U>0$. It is difficult to compute an order parameter for the antiferromagnetic ordering
because the ground states are inhomogeneous in real space and break the spin rotation symmetry. In principle, one could compute the square root of the staggered average of the correlation function~(\ref{eq:symcorr}) over all sites for long ladder lengths $L_x$ but the computational cost would be excessive with our DMRG program.

In summary, our results for the 3-2 ladder geometry agree
with the exact results in Refs.~\cite{lieb1989two} and \cite{Shen1994}. The ground state is
gapped for charge excitations and ferromagnetic with a total spin $S \neq \vert N_A - N_B \vert/2$. Moreover, the
electron spins are ferrimagnetically ordered.  Our results also quantify the precise
behavior of the correlation functions.

\begin{figure}
    \includegraphics[width=0.42\textwidth,angle=-90]{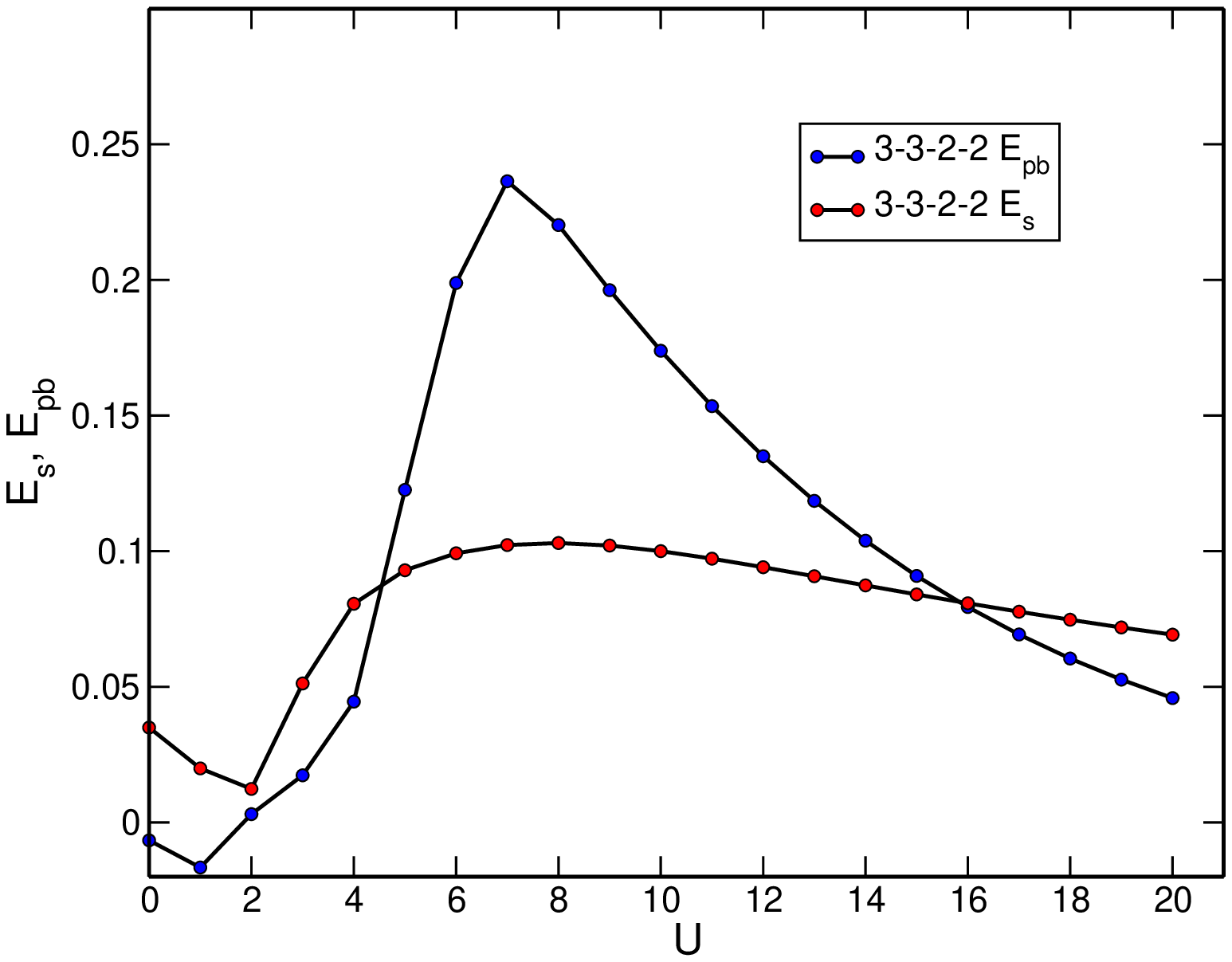}
    \caption{\label{6-4-Es-Epb-U} Spin gap $E_{s}$ and pair-binding $E_{pb}$ energies of the half-filled 3-3-2-2 ladder with $L_{x}=128$ as a function of the coupling strength $U$.} 
\end{figure}

\subsection{3-3-2-2 alternating rung geometry}

We now turn to the  Hubbard model on the 3-3-2-2 ladder geometry. 
This structure allows for unequal rung lengths, but unlike the 3-2 case,
this system represents a bipartite lattice with $N_A=N_B$. We have verified using DMRG that the ground state for $U>0$ at half filling has spin \(S=0\) and is not degenerate.

The charge gap $E_c$ increases with $U$ and is very close to the charge gap
of the 3-2 ladder geometry for $U \geq 5t$ as seen in Fig.~\ref{6-4-3-2-Ec-U} for a finite ladder length.
For weak coupling the charge gap of the 3-3-2-2 ladder is 
clearly smaller than in the 3-2 ladder.
For $U\geq 4t$ our DMRG data indicate that the charge gap of the 
3-3-2-2 ladder remains finite 
in the thermodynamic limit. 
For smaller $U$ we cannot determine whether the charge gap vanishes or is only very small in the thermodynamic limit.

\begin{figure}
        \includegraphics[width=0.5\textwidth]{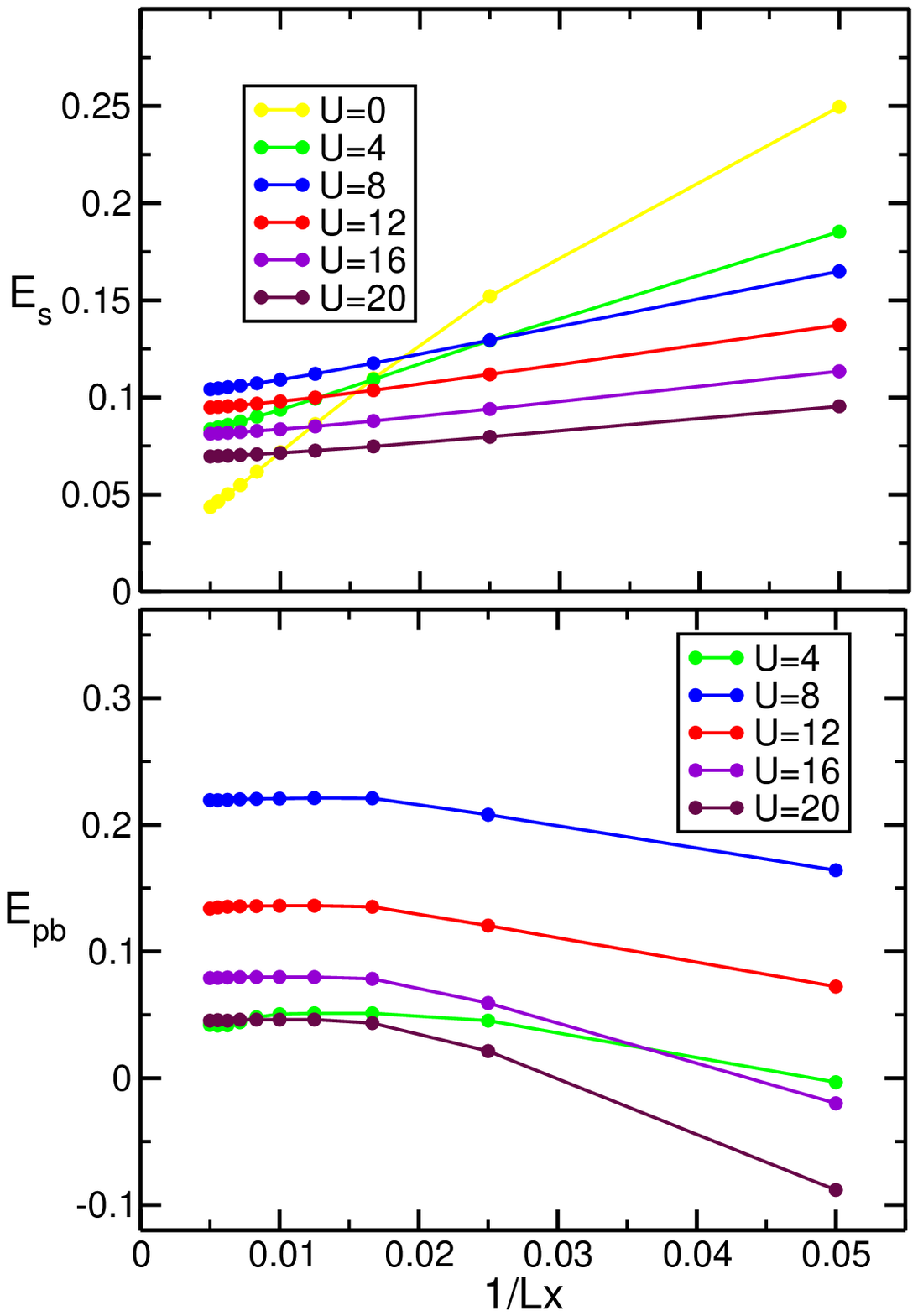}
        \caption{\label{6-4-Es-Epb}(a) Spin gap of the half-filled 3-3-2-2 ladder as a function of the inverse ladder length $1/L_x$ for different values of $U$.
       (b)  Pair-binding energy $E_{pb}$ of the half-filled 3-3-2-2 ladder as a function of $1/L_x$ for several values of $U$. 
}
\end{figure}

In contrast to the 3-2 system, spin excitations are gapped in the 3-3-2-2 ladder. 
As shown in Fig.~\ref{6-4-Es-Epb-U} for a ladder of finite length 
$L_x=128$, 
the spin gap $E_s$ increases with $U \geq 2$, reaches a maximum around $U=8$ and then decreases  $\sim t^2/U$ for strong coupling. 
This behavior is similar to the behavior found 
in homogeneous two-leg ladders~\cite{jeckelmann1998comparison}.
DMRG numerical errors and finite-size effects 
are not negligible for weak coupling $U < 2$ and are responsible for the nonmonotonic behavior of the spin gap.  (This is the case for many numerical
methods including Quantum Monte Carlo.  Larger $U$ breaks degeneracies
and eliminates finite size `shell effects' which are present at $U=0$.)
For $U\geq 4t$ we clearly see in Fig.~\ref{6-4-Es-Epb}(a)  that the spin gap converges to a finite value in the thermodynamic limit, in contrast to the noninteracting system.

\begin{figure}
        \includegraphics[width=0.4\textwidth,angle=-90]{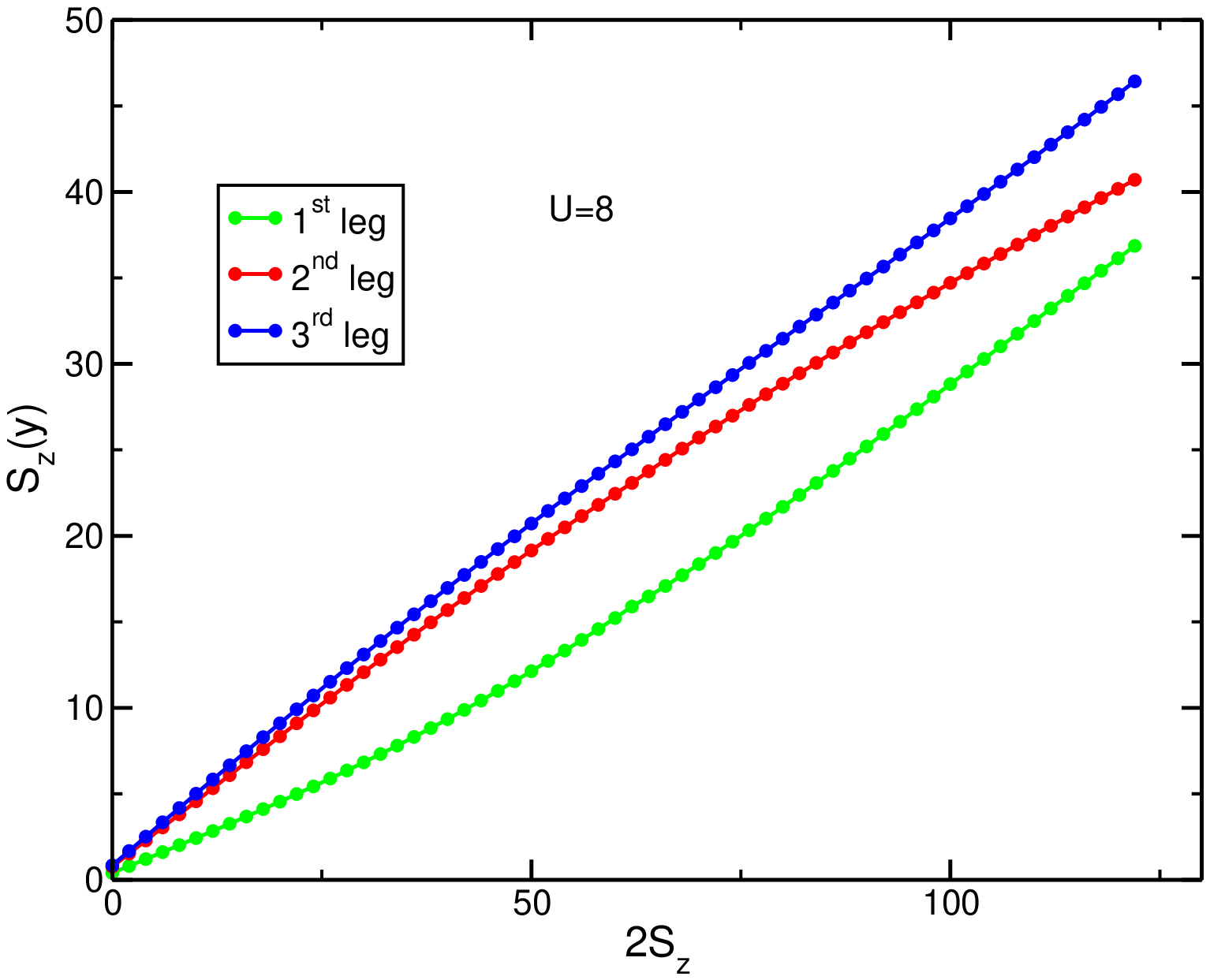}
        \caption{\label{total-Sz-4-6} Total spin density $S_z(y)$ on each leg 
        of the half-filled 3-3-2-2 ladder with $U=8$ and $L_x=128$ 
        as a function of the $z$-projection of the total spin.}
\end{figure}

In Fig.~\ref{total-Sz-4-6}  we show the total spin density on each leg of the 3-3-2-2 ladder with $U=8$ and $L_x=128$ as a function of 
the total spin $S_z$. The unpaired electron spins are distributed over the three legs but the first leg density is significantly lower than for the second and third legs.
Thus in the presence of an external magnetic field,
the unpaired electron spins are mostly localized on the second and third leg in the 3-3-2-2 system, in contrast
to the magnetization of the first leg in the 3-2 system,
see Fig.~\ref{3-2-GS-total-Sz-U8}(b).

The behavior of the single-particle gap $E_p$ is very similar to that of the charge gap $E_c$. In contrast to the case of the 3-2 ladder, however, we observe a difference between both gaps in the 3-3-2-2 ladder. The pair binding energy \cite{chakravarty1991electronic} is defined as
\begin{equation}
 E_{pb}=E_p -E_c.
\end{equation}
Finite positive values of $E_{pb}$ indicate that it is energetically preferable for electrons or holes injected into the half-filled band to form pairs~\cite{lin2007strong}. It has been shown for homogeneous 2-leg Hubbard ladders that the pair binding energy can be positive for some parameter ranges~\cite{coppersmith1989phase, chakravarty1991absence,jeckelmann1998comparison}.
The finite-size scaling shown in Fig.~\ref{6-4-Es-Epb}(b) confirms that the pair-binding energy is positive and remains finite in the thermodynamic limit of the 3-3-2-2 ladder at least in the regime $U>4t$.
As seen in Fig.~\ref{6-4-Es-Epb-U}, the pair binding energy
seems to reach its maximum $E_{pb} \approx 0.22t$  for  $U\approx 8$ where the spin gap $E_s$ is also the largest. 
Again this behavior is similar to the observation made 
for homogeneous symmetric~\cite{jeckelmann1998comparison} and anti-symmetric~\cite{Abdelwahab2015} two-leg Hubbard ladders.
One can conclude that the pair binding energy is intimately related to the behavior 
of the spin gap energy \cite{bennemann2011physics}.

\begin{figure}
        \includegraphics[width=0.4\textwidth, angle=-90]{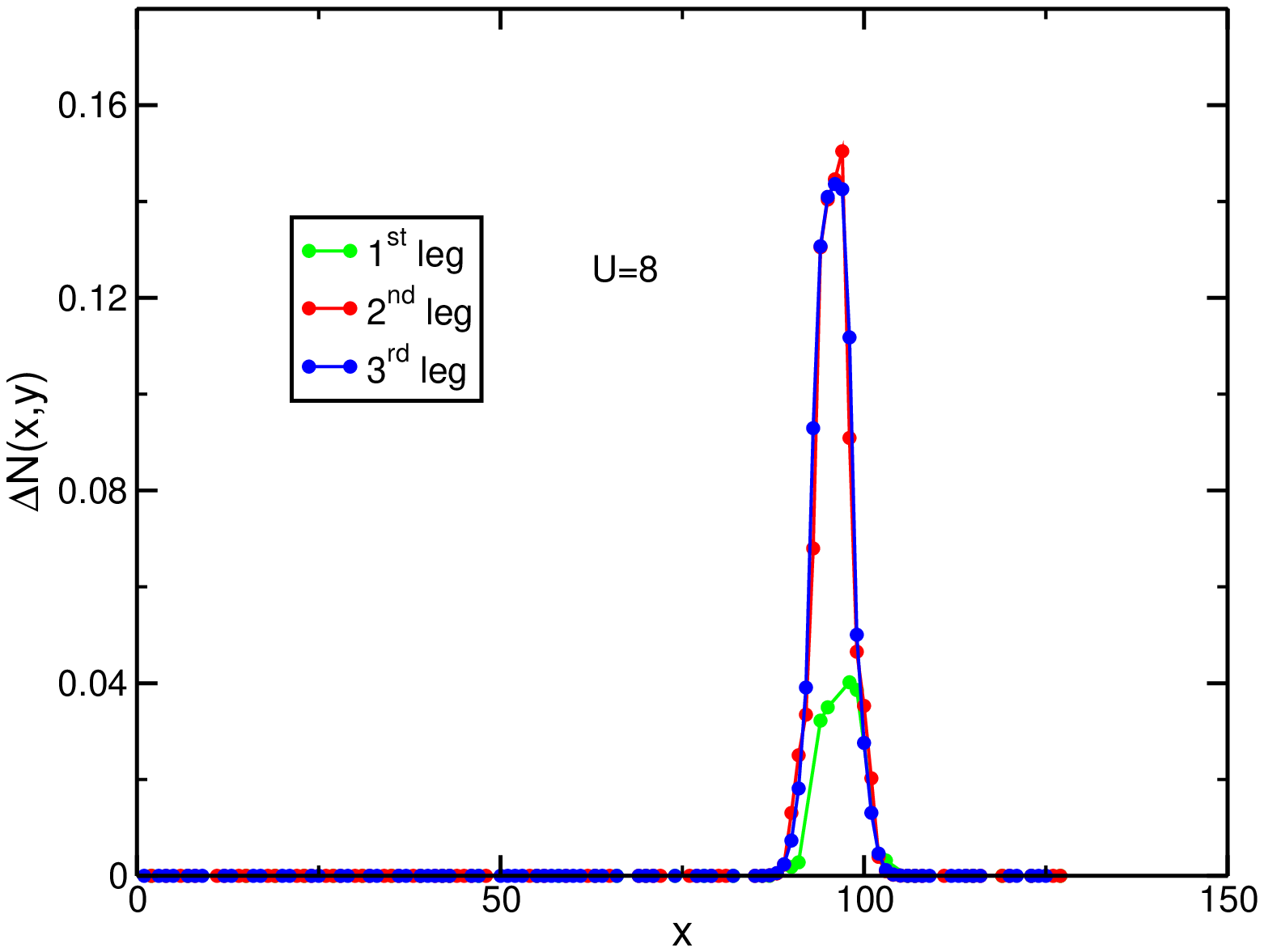}
        \caption{\label{6-4-charge-density-Sz-0} Change
        $\Delta N(x,y)$ in the charge density distribution $N(x,y)$ on the three legs ($y=1,2,3$) when two electrons are added to a  half-filled 3-3-2-2 ladder with $U=8$  and $L_x=128$.
        }
\end{figure}

As for the 3-2 system, the charge density $N(x,y)$ of the 
half-filled 3-3-2-2 ladder is uniformly distributed because of the particle-hole symmetry of the Hubbard Hamiltonian~(\ref{eq:Hfull}) on a bipartite lattice.
Figure \ref{6-4-charge-density-Sz-0} shows the change 
$\Delta N(x,y)$ of this charge density distribution
when two electrons are added to the half-filled system
 with $U=8$ and $L_x=128$. 
  As was the case for the 3-2 system, the increase of the charge density  in the first leg is smaller than the second and the third leg.
  The single peak resembles the distribution expected for a single particle in a box. This suggests that the two electrons build a pair (mostly localized on the second and third leg) in agreement with the observation of a finite binding energy. 
  The asymmetric distribution $\Delta N(x,y)$ is due to a poor DMRG convergence as already discussed in the previous section.
\begin{figure}
 \includegraphics[width=0.4\textwidth,angle=-90]{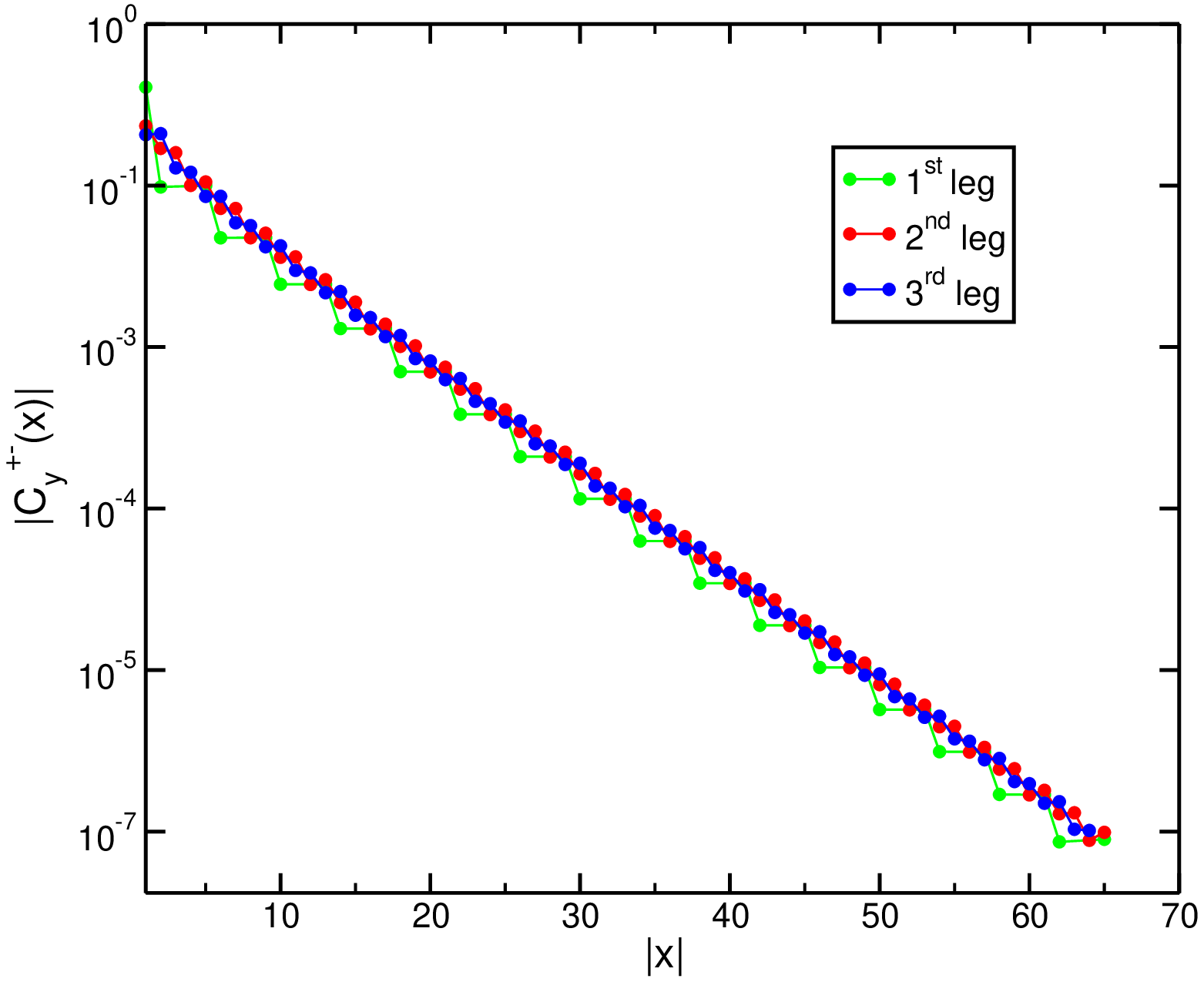}
\caption{\label{6-4-Spin-Correlation}Spin correlation function $C^{+-}_{y}(x)$ for the 3-3-2-2 geometry along each leg $y=1,2,3$ at half-filling with $U=8$ and $L_x=128$.}
\end{figure}

Spin correlations decrease exponentially with distance
in the half-filled 3-3-2-2 ladder, as expected for a spin gapped system. These correlations reveal short-range antiferromagnetic order.
Figure \ref{6-4-Spin-Correlation} illustrates
the exponential decrease of the correlation function
$C^{+-}_{y}(x)$ with the distance $x$ between two sites
on each leg $y$ for $U=8$ and $L_x=128$.
We see that the decay of spin correlations is
similarly fast in all legs.

\begin{figure}[b]
        \includegraphics[width=0.4\textwidth,angle=-90]{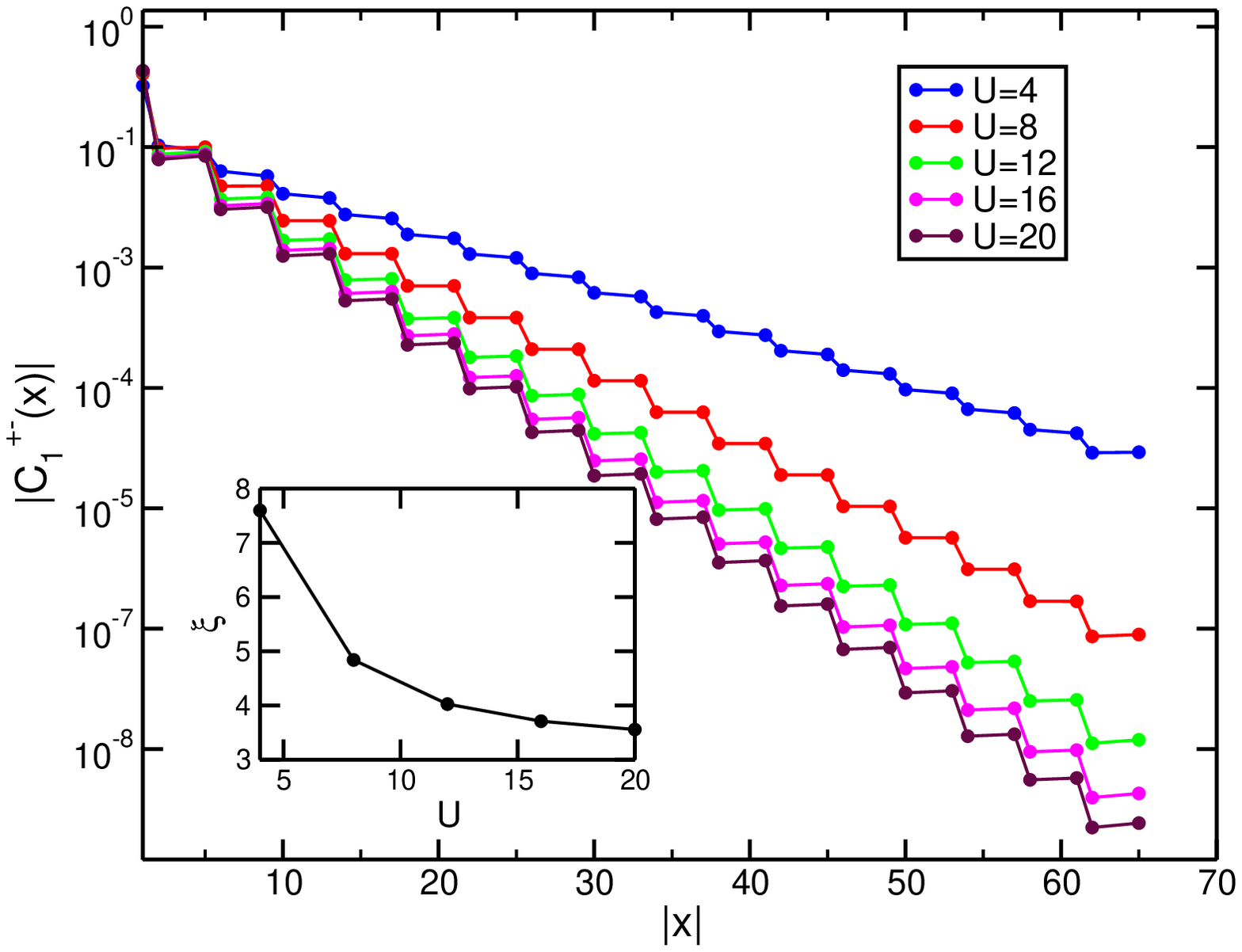}
        \caption{\label{Spin-Spin-Correlation-6-4-U-x} Spin correlation function $C^{+-}_{1}(x)$ in the first leg for the 3-3-2-2 geometry at half-filling for various values of $U$. The inset shows the derived correlation length $\xi$ as a function of $U$.}
\end{figure}

In Fig.~\ref{Spin-Spin-Correlation-6-4-U-x}, the exponential decay 
of spin correlations is plotted for different values of $U$. 
We derive a correlation length $\xi$ by fitting these data to an exponential function. The results are shown in the inset of Fig.~\ref{Spin-Spin-Correlation-6-4-U-x}
as a function of $U$. The correlation length decreases rapidly with increasing coupling $U$ but seems to saturate at a finite value for strong coupling. This can be explained by the fact that the correlation length is roughly given by the ratio between the band width and the gap of spin excitations. For weak $U$ we know that
the band width is $\propto t$ while the spin gap is small,
see Fig.~\ref{6-4-Es-Epb-U}, leading to a divergence of $\xi$ for $U\rightarrow 0$.
For strong $U$, however, both the band width and the spin gap scale with the effective exchange coupling
$J \propto t^2/U$.

In order to get information about the relative orientation of spins on nearest-neighbor sites,
we calculate the spin bond order defined as,
\begin{equation}\label{eq:spinbondorder}
 B(x,y,x',y') = \langle  (n_{x,y,\uparrow}-n_{x,y,\downarrow})
( n_{x',y',\uparrow}-n_{x',y',\downarrow} )
 \rangle
\end{equation}
where $(x,y)$ and $(x',y')$ are nearest-neighbor sites.

Note that  the isotropic spin bond order is equal to~(\ref{eq:spinbondorder}) up to a factor $3/4$  because the ground state of the 3-3-2-2 ladder geometry
has spin $S=0$.

\begin{figure}
         \includegraphics[width=0.4\textwidth, angle=-90]{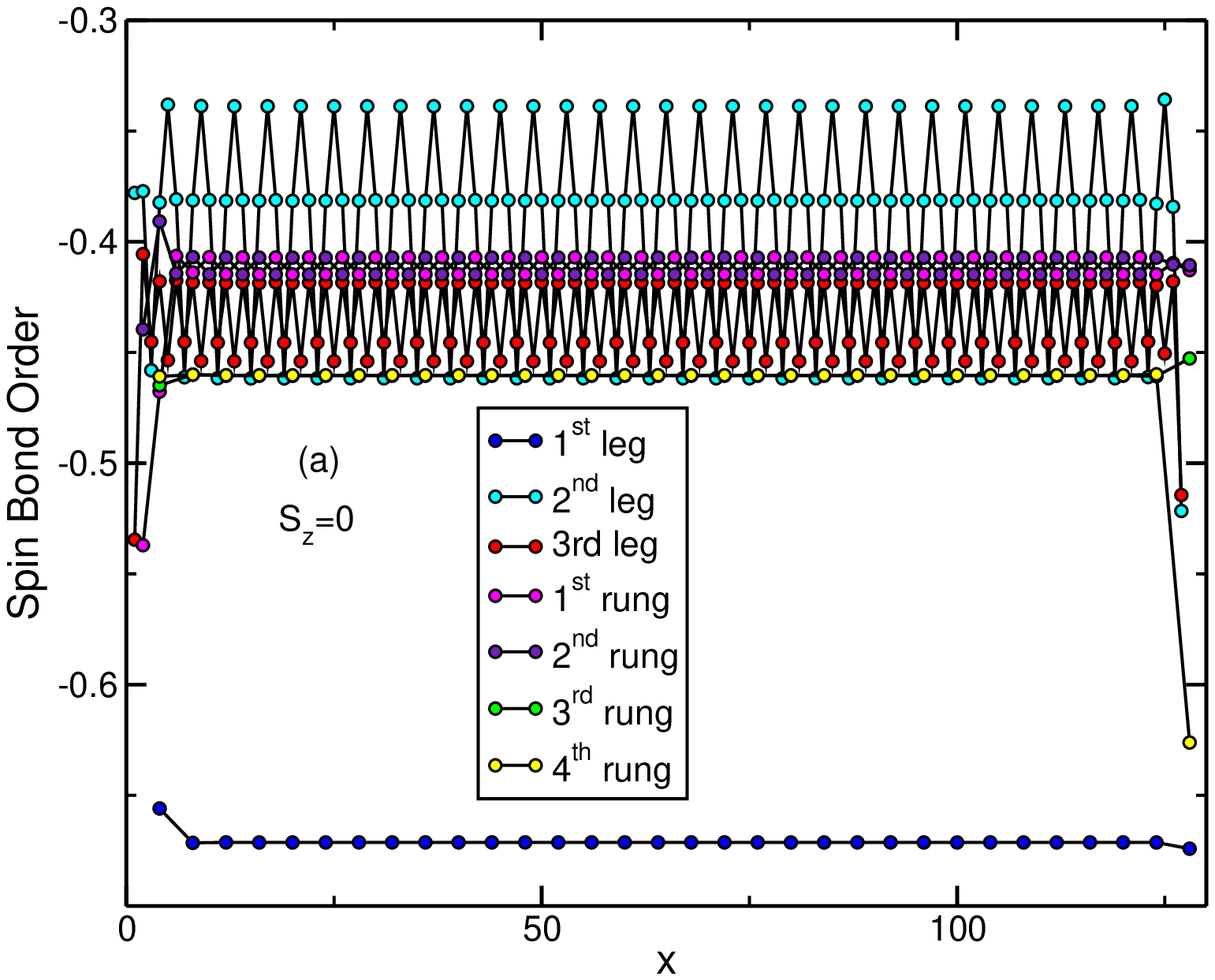}
        \includegraphics[width=0.4\textwidth]{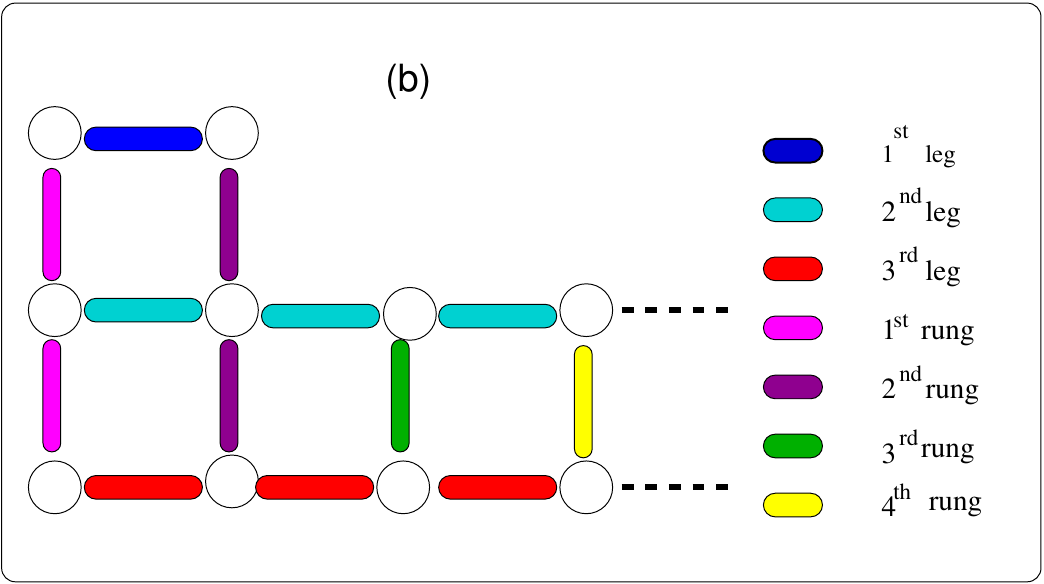}
        \caption{\label{6-4-spin-bond-order}(a) Spin bond order between nearest-neighbor sites for the 3-3-2-2 ladder at half-filling with $U=8$ and $L_x=128$. (b) Schematic illustration of the 3-3-2-2 configuration. 
          }
\end{figure}

Figure \ref{6-4-spin-bond-order} shows DMRG results for this spin bond order.
We see that the spin bond is much stronger between 
the nearest-neighbor spins on the first leg than
all other spin bonds.
This suggests that each nearest-neighbor pair on the first leg builds a strong singlet which is then weakly coupled to the
rest of the system.  The second and third legs build
an effective two-leg ladder. 
This structure is illustrated in Fig.~\ref{6-4-spin-bond-order}(b).
This (partial) decoupling of the first leg from the other two legs explains the similarities
between the 3-3-2-2 system and two-leg ladders that we observe, in particular the pair binding of two added electrons. 

\section{Conclusion}

The alternation between gapped and ungapped spectra in uniform ladder systems
has been of profound importance in condensed matter physics
\cite{ajiro89,azuma94,regnault94,white1994resonating,granroth96,takigawa96,dagotto1996surprises,greven96}.
The focus of the present work was to explore spin and charge properties of novel Hubbard ladder 
systems whose geometry consists of alternating numbers of sites per rung and
thus `mix' the structures which are associated with these two situations.
Analogs of such structures, chains with spins alternating between
$S=1/2$ and $S=1$, have revealed a number of unexpected phenomena driven by a competition
between half-integer and integer spin physics,
including transitions between antiferromagnetic
and ferromagnetic behavior as the temperature is varied
\cite{pati97,yamamoto98a,yamamoto98b,langari00}.
Here we can investigate such physics for {\it itinerant} electrons
rather than quantum spins,
in geometries which
combine features of odd and even rung systems. 
Our consideration of the Hubbard rather than the Heisenberg Hamiltonian
also allows us to consider the effects of the unique zero energy
bands associated with bipartite lattices
with unequal number of sites per sublattice. 
Our findings are consistent with available analytical results
for such  geometries
~\cite{lieb1989two,Shen1994}.

Investigating two different geometries  has allowed us to isolate the effects of unequal sublattice site numbers from that of a periodically varying number of sites per rung.
We have found that the low-energy properties of the two three-leg ladder systems at half filling differ radically. The magnetic properties of an alternating ladder with unequal number of sites in each sublattice
are similar to those predicted for the Hubbard model on a Lieb lattice. The ground state has a total spin $S$ proportional to the system size and the electron spins are ordered ferrimagnetically. Thus this is a rare example of long-range magnetic order in a one-dimensional quantum system with short-range interactions. Moreover, we have found that electron or hole pairs added to the half filled system do not seem to bind.
In contrast, the properties of the alternating ladder with equal number of sites in each sublattice resemble the properties found in two-leg ladders because the spins on the depleted leg tends to build strong singlets. The ground state is paramagnetic and non-degenerate. Added electron and hole pairs tend to bind
with a binding energy that seems to be set by the size
of the spin gap.

Past investigations of uniform spin and fermion ladders already revealed
profoundly different low energy magnetic properties, e.g. depending on the number of 
legs~\cite{giamarchi2004quantum}.  Attention subsequently turned to refinements including 
DMRG calculations that revealed transitions between gapped paramagnetic
and ferrimagnetic phases in two-leg ladders with
alternating spin-1/2 and spin-1 degrees of freedom~\cite{pati97,ivanov98,fukui98,chandra10}.
Our work shows that ladders with varying number of sites per rung exhibit a similar rich physics, 
including long-range magnetic order, while being amenable to well-established methods for one-dimensional quantum many-body systems. 

It will be interesting to explore further aspects of the physics of uniform ladders
in our alternating geometry, including
both how pairing correlations decay \cite{dagotto92,sigrist94},
and also the nature of spin correlations
in the vicinity of magnetic impurities\cite{wang99}.

\acknowledgments

The work of R.S.~was supported by the 
grant DE‐SC0014671 funded by
the U.S. Department of Energy, Office of Science.

\appendix
%
\section{Homogeneous 3-leg ladder}\label{Homogeneous3legladder}

\begin{figure}
        \includegraphics[width=0.4\textwidth]{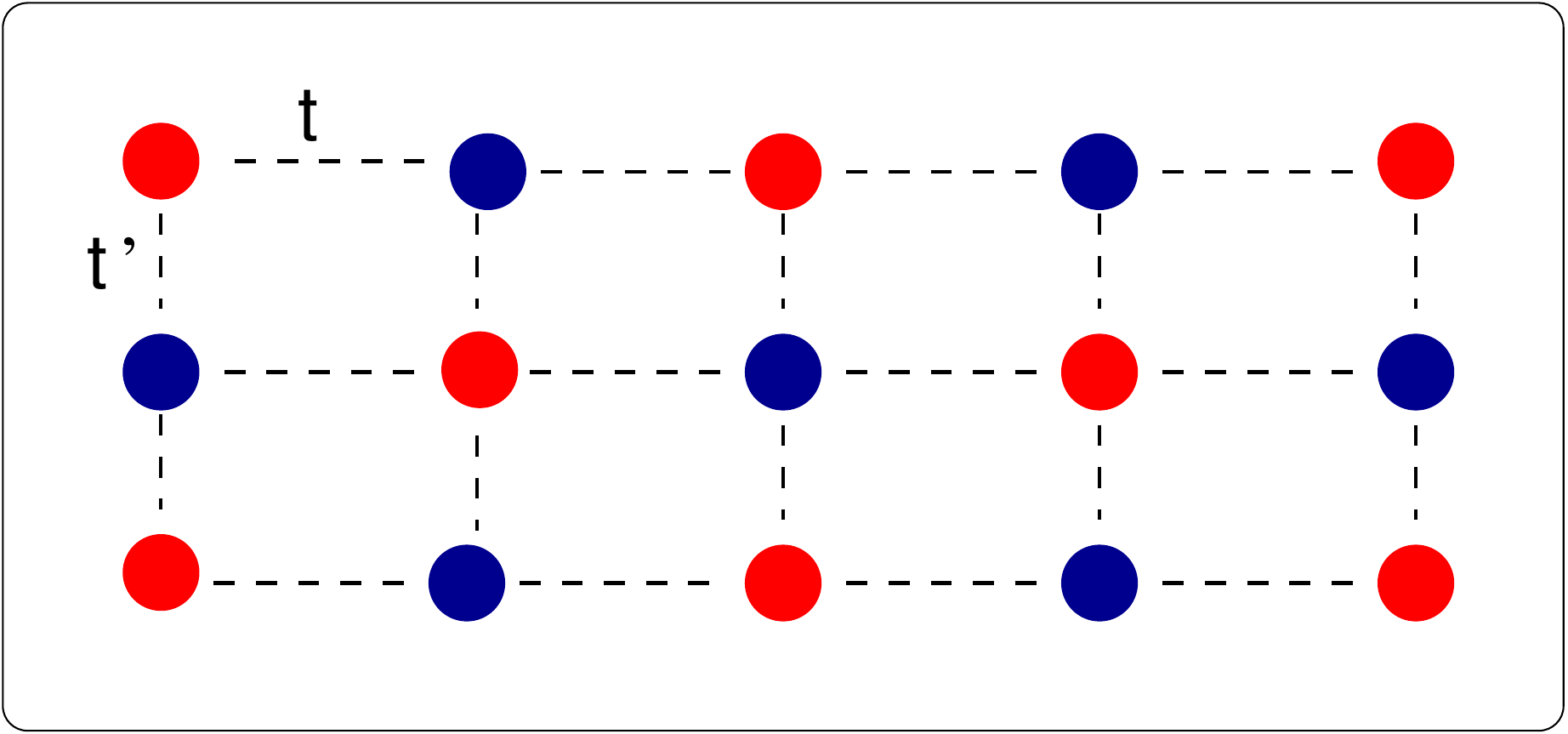}
        \caption{\label{3legladder}  Homogeneous 3-leg ladder with intra-leg hopping term $t$ and inter-leg hopping term $t^{\prime}$.}
        \label{3-homogeneous-legs}
\end{figure}

Here we review some results for homogeneous noninteracting 3-leg ladders
and detail the method that we also use to study the alternating ladder geometries
in Sec.~\ref{AlternatingGeometries}.
The structure of the homogeneous 3-leg ladder lattice is shown in Fig~\ref{3legladder}. It consists of three chains, with intra-chain hopping $t$ and inter-chain hopping $t^\prime$. We focus on the band structure and DOS of this system with $ t=t^{\prime}=1$ for both noninteracting and interacting cases.

If we use periodic boundary conditions in the leg-direction and the non-interacting Hamiltonian~(\ref{3legHubbardU0})
is periodic with period $d$ in that direction, we can write it as a sum 
\begin{equation}
H_0 = \sum_k H_k
\end{equation}
of commuting many-body operators $H_k$. Each $H_k$  acts only 
on the single-particle Bloch states with the wave number 
\begin{equation}\label{wavenumber}
k = \frac{2\pi j d}{L_x}
\end{equation}
in the first Brillouin zone
where the quantum number $j$ satisfies $ -L_x/(2d) < j \leq L_x/(2d)$. $L_x$ is the number of rungs and 
 $N_c=L_x/d$ is the number of unit cells or equivalently the number of wave numbers $k$.
 The Bloch states are given by the 
 transformation 
\begin{equation}\label{BlochTransform}
 d^{\dag}_{k,x,y,\sigma} = \sqrt{\frac{d}{L_x}}\sum_{n=1}^{L_x/d} c^{\dag}_{x+nd,y,\sigma} \exp(ikn).
\end{equation}
where $x=1,\dots,d$.
For the homogeneous 3-leg ladder we have $d=1$.
Each many-body Hamiltonian $H_k$ acts on as many single-particle states as there are sites in one unit cell.
For the homogeneous 3-leg ladder this dimension is three. Thus we obtain
the $3\times 3$ matrix representation
of $H_k$ for single-particle states
 \begin{equation}\label{3legMomentumSpace}
H^{(1)}_{k} = 
\begin{pmatrix}
   -2t \cos(k) &  -t^\prime   &  0 \\
  -t^\prime  & -2t \cos(k) &-t^\prime \\
   0  &  -t^\prime &  -2t \cos(k)  &
\end{pmatrix}.
\end{equation}

\begin{figure}[ht]
       \includegraphics[width=0.48\textwidth]{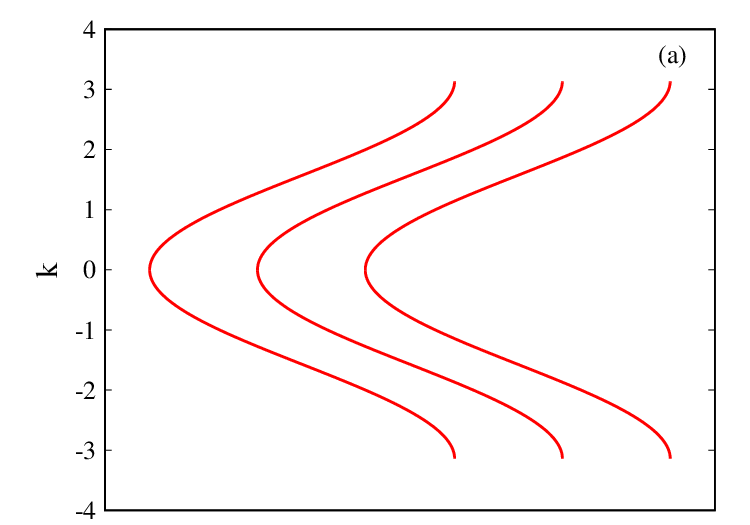}
       \includegraphics[width=0.48\textwidth]{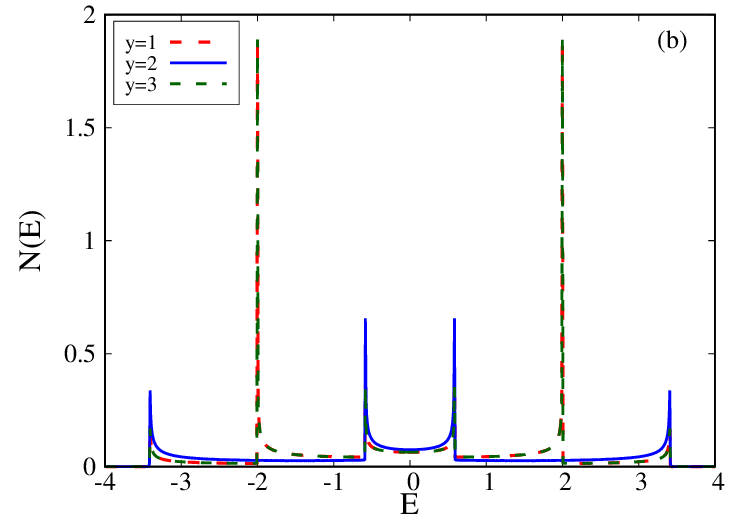}       
               \caption{ \label{EB-DOS-3-leg-ladder} (a) Band structure for the noninteracting homogeneous 3-leg ladder with $t=t^\prime=1$. (b) Corresponding density of states on each leg $y=1,2,3$. Note that $N(E,1)=N(E,3)$. }
\end{figure}

In order to gain some insight into this system, we calculate the single-particle eigenenergies by diagonalizing this matrix, which leads to three  bands with the dispersion relation
\begin{equation}
E_{k,b}= -2t\cos(k)+\varepsilon_{b} 
\label{dispersion}
\end{equation}
where $\varepsilon_{b}=0,\pm \sqrt{2}t^{\prime}$. The index $b \ (=1,2,3)$ numbers the bands. The band structure is shown in Fig.~\ref{EB-DOS-3-leg-ladder} (a).
As we consider a homogeneous ladder with inequivalent legs, it is useful to calculate the leg-resolved DOS
\begin{equation}
 N(E,y) = \frac{1}{N_c} \sum_{{k,b}} \vert\psi_{k,b}(y)\vert^{2} \delta{(E-E_{k,b})}
 \label{DOS_one_leg}
\end{equation}
for $y=1,2,3$
rather than the total DOS~(\ref{DOS}).
Here $\psi_{k,b}(y)$ represents the eigenvector of the matrix (\ref{3legMomentumSpace}) corresponding to the 
eigenenergy $E_{k, b}$ and $N_c=L_x$. 
The DOS for the homogeneous 3-leg ladder 
is plotted in Fig.~\ref{EB-DOS-3-leg-ladder}(b). It shows six Van Hove singularity peaks for the first and third leg, with $N(E,1)=N(E,3)$. The DOS on the second leg exhibits
only four singularities because the band~(\ref{dispersion}) with $\varepsilon_b=0$  is anti-symmetric under reflection
in the $y$-direction and thus the corresponding single-particle eigenstates vanish on the middle leg, 
$\psi_{k,b}(x,y=2)=0$. 
For all finite values of $t$ and $t^{\prime}$ the system is metallic, i.e. there is at least one band crossing the Fermi level
$E_F=0$ at half filling.

\section{Hamiltonian matrix for the noninteracting 3-3-2-2 ladder geometry}
\label{app:matrix}
The single-particle matrix representation  $H^{(1)}_k$ of the Hamiltonians $H_k$ for the noninteracting 3-3-2-2 ladder geometry is the following  $10 \times 10$ matrix with $\widetilde t_j =t_j\exp(ik)$.
\begin{widetext}
\begin{align*}
 H^{(1)}_k =  
\begin{pmatrix}
0& -t_{1}& 0& 0& 0& -t_{5}& 0& 0& 0& \widetilde t_{15}\\
-t_{1}& 0& -t_{2}& 0& -t_{4}& 0& 0& 0&  \widetilde t_{14}& 0\\
0& -t_{2}& 0&-t_{3}& 0& 0& 0& 0& 0& 0\\
0& 0& -t_{3}& 0& -t_{6}& 0& 0& 0& 0& 0\\
0& -t_{4}& 0& -t_{6}& 0& -t_{7}& 0& -t_{8}& 0& 0\\
-t_{5}& 0& 0& 0& -t_{7}& 0&-t_{9}& 0& 0& 0\\
0& 0& 0& 0& 0& -t_{9}& 0& -t_{10}& 0& -t_{12}\\
0& 0& 0& 0& -t_{8}& 0& -t_{10}& 0& -t_{11}& 0\\
0&-\widetilde t^{*}_{14}& 0& 0& 0& 0& 0&-t_{11}& 0& -t_{13}\\
-\widetilde t^{*}_{15}& 0& 0& 0& 0& 0&-t_{12}& 0& -t_{13}& 0\\
\end{pmatrix}
\end{align*}
\end{widetext}

%
\bibliographystyle{unsrtnat}

\bibliography{bibliography_alternation}

\end{document}